# Crossover from quantum to classical transport

Dirk K. Morr

*Department of Physics, University of Illinois at Chicago, Chicago, IL 60607, USA*

e-mail: dkmorr@uic.edu

# Crossover from quantum to classical transport

Understanding the crossover from quantum to classical transport phenomena has become of fundamental importance not only for technological applications due to the creation of sub-10nm transistors – an important building block of our modern life – but also for elucidating the role played by quantum mechanics in the evolutionary fitness of biological complexes. This article provides a basic introduction into the nature of charge and energy transport in the quantum and classical regimes. It discusses the characteristic transport properties in both limits, and demonstrates how they can be connected through the loss of quantum mechanical coherence. The salient features of the crossover physics are identified, and their importance in opening new transport regimes and in understanding efficient and robust energy transport in biological complexes are demonstrated.

## 1. Introduction

Our daily life experiences are based on classical physics: the way we drive a car, we use electricity to cook, or pick up a book, is governed either by Netwon's laws of motion[1], or by Maxwell's equations of electrodynamics [2]. On the other hand, everything that surrounds us is made of atoms, neutrons, protons, electrons and quarks, whose behaviour and properties are determined by the laws of quantum mechanics [3]. Many of those laws, however, do not have a "clsssical analog" and would appear rather foreign if transferred into our classical world: the uncertainty principle [4], for example, is clearly something that if it existed in our classical world, would make walking to school or throwing a baseball rather difficult. The question therefore immediately arises of how the laws of quantum mechanics which govern the atomic and sub-atomic world evolve into the laws of classical physics in our macroscopic world. This question has become of particular importance in those instances where the quantum and classical worlds have started to overlap. While the flow of an electrical charge current through a piece of copper, for example, is determined by Maxwell's classical equations, the flow of currents through a macroscopic superconductor [5] is governed by quantum mechanics. These two worlds approach each other in the chips embedded in the smart phones, tablets and computers we use on a daily basis. Moore's law [6] has reliably predicted for the last 50 years that the density of transistors (which are the basic building blocks for any computational technology) on an integrated chip doubles every two years. Since computational processes require the flow of charge currents, it follows that charges are flowing through increasingly smaller and smaller systems, with sizes of transistors having reached the sub-10nm scale in the last few years. While the flow of charge in macroscopic leads (let us say that those with diameters of millimetres or larger) are governed by Ohm's law and Maxwell's equations, there surely is a crossover range, presumably in the tens to hundred nanometer range, below which it is the laws of quantum mechanics that determine the flow of charges [7]. This raises two interesting questions. First, what sets the length scale for the crossover from classical to quantum mechanical transport? Second, what is the correct (theoretical) starting point to describe this crossover? These two questions are not only of fundamental scientific interest, but also of great importance for technological applications. Interestingly enough, similar questions have also attracted considerable interest over the last few years in the field of light-harvesting, photosynthetic biological complexes [8]. In these systems, energy from the sunlight is converted into an exciton – a particle-hole pair – which is then transported and ultimately stored in a "battery". While such biological complexes were believed to be inherently classical since they represent open systems, and thus strongly interact with the outside world, a series of experiments [9] over the last ten years has provided strong evidence that the nature of energy transport in biological systems might actually lie in the crossover region between the classical and quantum mechanical worlds.

To address the first of the two questions raised above, we need to understand the nature of classical and quantum charge transport. When a current flows through a simple conductor, such as a piece of copper, the electrons which make up the current, are constantly scattered, which gives rise to the resistivity of the material. There are three different mechanisms by which electrons are scattered: scattering off the vibrations of the lattice (i.e., phonons), scattering by other electrons, which arises from the Coulomb repulsion between electrons, and scattering off (static) defects or impurities in the material's lattice structure. Since scattering by electrons or phonons results not only in the exchange of momentum but also of energy, these processes are inelastic in nature and lead to the destruction of *coherence*, one of the most fundamental concepts underlying quantum mechanics.

Coherence manifest itself in the well-defined time evolution of wave-functions, which is represented by a phase factor $e^{iEt/\hbar}$ where $E$ is the energy of the wave-function. However, in an inelastic scattering event, the energetic state of the electron is changed, and any information about its phase (and thus coherence) is lost. The extent of the scattering is characterized by an inelastic scattering time (or lifetime) $\tau$ which is the time between scattering events, and a corresponding inelastic mean-free path, $l = v_F \tau$, where $v_F$ is the Fermi velocity of the electrons. If an electron can traverse a system without being inelastically scattered – implying that $l$ is larger than the systems size $L$ -- then quantum mechanical coherence can be established across the system, and its properties are governed by the laws of quantum mechanics. In this case, the transport is considered to be *ballistic*. On the other hand, for $l \ll L$, the frequent inelastic scattering of electrons while moving through the system completely destroys quantum mechanical coherence, leading to diffusive and thus classical transport. As a result, the transport properties of the system are described by the classical laws of Ohm and Kirchhoff. The crossover between these two regimes occurs for $l \sim L$. Interestingly enough, when $l \lesssim L$, the local transport properties, such as the spatial form of the charge flow, still reflects the properties of ballistic transport, while the global transport properties are already classical in nature [10]. Since an increase in the strength of the electron-electron or electron phonon scattering leads to smaller mean free paths, the transition from quantum mechanical behaviour for $l > L$ to classical behaviour for $l \ll L$ can be achieved either by increasing the system size, or by increasing the strength of the electronic scattering, as schematically shown in Fig.1.

On the other hand, in the scattering of electrons off (static) defects, such as imperfections of the lattice, only momentum, but no energy is exchanged, and the scattering process is therefore elastic. This mechanism persists down to zero temperature, and gives rise to the residual resistivity of metals -- a measure of their purity -- which is characterized by an *elastic* mean-free path. In fully quantum mechanical (coherent) systems, the elastic scattering off defects can give rise to the localization of electronic states, first predicted by Anderson [11]. This localization is accompanied by a suppression (in the case of weak localization [12]) or vanishing (in the case of strong or Anderson localization [11]) of the system's conductance. Thus inelastic scattering gives rise to the crossover from quantum to classical transport, while elastic scattering in quantum system give rise to the evolution from delocalized ballistic transport to weak or strong localization.

This brings us to the second question: what is the correct (theoretical) starting point to describe this crossover? To answer this question, we need to consider a fundamental difference in the nature of particles between the quantum mechanical and classical level. In quantum mechanics the concept of particle-wave duality reflects the observation that particles can possess particle-like but also wave-like properties. The wave-like nature of a particle can be observed as diffraction patterns in two-slit experiments, or can be directly made visible as standing waves of electrons in confined geometries, such as quantum corrals on the surfaces of metals [13, 14]. Furthermore, the ability of waves to interfere gives rise to fundamentally new phenomena related to transport, such as Anderson or weak localization of wave-functions [11, 12, 15-17], the shot-noise related to the flow of charges [18-21] or the spatial form of current patterns [22-24]. All of these phenomena, however, are unknown in the classical world where particles do not possess wave-like features. These observations suggest that any theory describing the cross-over from quantum mechanical to classical transport needs to start from a fully quantum mechanical description of the particles taking part in the transport, and identify those mechanisms by which the properties of these particles evolve into those of classical particles. This article reviews some of the recent progress that has been made in achieving this goal.

The rest of the article is organized as follows: in Sec. 2, I will briefly review the most salient concepts of the non-equilibrium Keldysh formalism necessary to understand and describe transport properties in the quantum limit. This section can be skipped by the reader, and simply referred back to as necessary. In Sec. 3, I discuss the transport properties characteristic of classical systems, before reviewing in Sec. 4 the transport properties of quantum systems. In Sec. 5, I demonstrate how the crossover between the quantum and classical limits which are characterized by qualitatively different transport properties can be described. In Sec.6, I review recent progress in imaging the flow of charge at the nanoscale. In Sec.7, I discuss crossover phenomena that emerge in the energy transport of biological complexes, before presenting the conclusions in Sec. 8.

## 2. Theoretical Formalism: From quantum mechanical ballistic to classical transport in Nanoscopic and Mesoscopic Systems

In the following, I briefly review the most salient features of the non-equilibrium Keldysh formalism for the study of charge and energy transport. The reader is referred to some excellent reviews on other theoretical approaches to transport, such as the quasi-classical Boltzmann theory [21] or the random matrix theory [25] for the study of mesoscopic systems.

### *2.1 General Formalism and the quantum mechanical ballistic limit*

The above discussion has shown that the crossover from classical to quantum transport has become of particular interest in nanoscale or mesoscale systems. I will therefore focus our discussion on systems of this length scale and begin by briefly discussing a theoretical framework, based on the non-equilibrium Keldysh Green's function formalism [26-29], that can not only describe the transport properties of quantum mechanical systems, but also the evolution of these properties from the quantum to the classical limit. This formalism is ideally suited to consider the flow of charge or energy in systems ranging from simple metals [22, 23], graphene [30, 31] or topological insulators [32] to networks of quantum dots and biological complexes [33]. It allows to systematically investigate the effects of lattice imperfections, of impurities, or of electron-phonon interactions on the system's transport properties order by order in the interaction strength.

Starting point for this investigation is the description of a system in terms of a connected network, as schematically shown in Fig.2(a). Here, the nodes or sites (gray dots) can represent single atoms, molecules, or quantum dots, while the links connecting the sites represent allowed paths along which electrons or excitons can move. The networks are connected to narrow and/or wide leads, which represent the source and sink through which electrons or excitons can enter or exit the network. For simplicity, we restrict the following discussion to (a) systems of square-lattice geometry, and (b) charge transport; the transport of excitons and other lattice types, such as hexagonal lattice, or topological insulators, will be discussed later.

A network of square lattice geometry in which sites represent individual atoms with given electronic energy levels (we will point out below, which modifications are necessary to consider molecules or quantum dots) and electrons can only hop between nearest-neighbour sites, is described by the Hamiltonian

$$H = -t \sum_{<\mathbf{r},\mathbf{r'}>,\sigma} c^\dagger_{\mathbf{r},\sigma} c_{\mathbf{r'},\sigma} - E_0 \sum_{\mathbf{r},\sigma} n_{\mathbf{r},\sigma} - t_l \sum_{\mathbf{l},\mathbf{L}_i,\sigma} \left( d^\dagger_{\mathbf{l},\sigma} c_{\mathbf{L}_i,\sigma} + h.c. \right)$$

$$- t_l \sum_{\mathbf{r},\mathbf{R}_i,\sigma} \left( d^\dagger_{\mathbf{r},\sigma} c_{\mathbf{R}_i,\sigma} + h.c. \right) + H_{lead} \quad . \tag{1}$$

Here the first term on the right-hand-side represents the hopping of electrons between nearest-neighbour sites in the network [described as grey links in Fig.2(a)], with $-t$ being the hopping amplitude, and $c^\dagger_{\mathbf{r},\sigma}, c_{\mathbf{r},\sigma}$ being the fermionic creation and annihilation operators that create or annihilate an electron with spin $\sigma$ at site $\mathbf{r}$. The second term represents the on-site energy of an electron at site $\mathbf{r}$, where $n_{\mathbf{r},\sigma} = c^\dagger_{\mathbf{r},\sigma} c_{\mathbf{r},\sigma}$ is the particle number operator. The third and fourth terms describe the hopping between the network and the left and right leads, respectively, with $\mathbf{L}_i$ and $\mathbf{R}_i$ denoting the sites in the network that are connected to the leads. Finally, $H_{lead}$ describes the electronic structure of the leads. We assume below that the hopping between the sites is sufficiently large (i.e., the network is a good metal) that effects arising from Coulomb repulsion, such as the Coulomb blockade in quantum dots [34-36], can be neglected.

To discuss the transport properties of the network, we first need to obtain its electronic structure by solving the Schrödinger equation using the above Hamiltonian

$$H|\psi_{\mathbf{k}}\rangle = E_{\mathbf{k}}|\psi_{\mathbf{k}}\rangle \qquad (2)$$

with $E_{\mathbf{k}}$ being the energy eigenvalues, and $|\psi_{\mathbf{k}}\rangle$ being the corresponding eigenstates, described by the quantum number $\mathbf{k}$.

In the absence of leads, the network is closed, and therefore possesses a discrete set of energy eigenstates which can be obtained by solving the quantum mechanical particle-in-a-box problem [22, 24]. For the network shown in Fig.2(a) the spatially dependent wave-functions take the form

$$\psi_{\mathbf{k}}(\mathbf{r}) = \langle \mathbf{r}|\psi_{\mathbf{k}}\rangle = \frac{2}{\sqrt{(N_x + 1)(N_y + 1)}} \sin(k_x r_x)\sin(k_y r_y) \qquad (3)$$

where $N_{x,y}$ is the number of rows and columns of the network, $\mathbf{k} = (k_x, k_y)$ is the Bloch wave-vector of a given eigenstate, and the prefactor $2/\sqrt{(N_x + 1)(N_y + 1)}$ arises from the normalization of the wave-function. The Bloch wave-vectors $k_{x,y}$ are determined by the requirement that the wave-function vanish outside of the network, i.e. at $r_{x,y} = 0$ or $r_{x,y} = (N_x + 1)a_0$, which yields $k_i = n_i\pi/(N_i + 1)a_0$ $(i = x, y)$ with $n_i = 1, ..., N_i$ and $a_0$ is the lattice constant. Thus, the wave-vectors and the corresponding eigenenergies

$$E_{\mathbf{k}} = -2t[\cos(k_x a_0) + \cos(k_x a_0)] - E_0 \qquad (4)$$

of the wave-functions are discrete.

The electronic structure and transport properties of such a network are best discussed in terms of Green's functions [37]. Of particular importance are the retarded and advanced Green's functions which can be obtained from the above eigenenergies and wave-functions via

$$G^{r,a}(\mathbf{r}, \mathbf{r}', E) = \sum_{\mathbf{k}} \frac{\psi_{\mathbf{k}}^*(\mathbf{r})\psi_{\mathbf{k}}(\mathbf{r}')}{E - E_{\mathbf{k}} \pm i\delta} \qquad (5)$$

where $\delta = 0^+$ is an infinitesimally small and positive number. Here the superscripts $r$ and $a$ denote the retarded and advanced Green's functions, respectively, and $\mathbf{r}$ and $\mathbf{r}'$ denote two sites in the network. Quite generally, Green's functions represent the electronic correlations of a system, and can be considered a measure for the quantum mechanical amplitude of a process in which an electron propagates from site $\mathbf{r}$ to site $\mathbf{r}'$. An important physical quantity that describes the electronic structure of a network is the local density of states (LDOS); it is obtained from the retarded local Green's function with $\mathbf{r}' = \mathbf{r}$ via

$$N(\mathbf{r}, E) = -\frac{1}{\pi} \text{Im}\left[G^r(\mathbf{r}, \mathbf{r}, E)\right] = \sum_{\mathbf{k}} |\psi_{\mathbf{k}}(\mathbf{r})|^2 \delta(E - E_{\mathbf{k}}) . \qquad (6)$$

Since the energy eigenstates of the network appear as $\delta$-functional peaks in the LDOS at $E = E_{\mathbf{k}}$ with spectral weight $|\psi_{\mathbf{k}}(\mathbf{r})|^2$ at site $\mathbf{r}$, knowledge of the LDOS provides not only insight into the energy of the network's eigenstates, but also into the modulus of their spatially dependent wave-function.

Green's functions are ideally suited to understand and describe charge transport in a network. In particular, within the non-equilibrium Keldysh Green's function formalism [27-29], the current between sites $\mathbf{r}$ and $\mathbf{r}'$ in a network is given by [38]

$$I_{\mathbf{rr}'} = -2\frac{e}{\hbar}g_e \int_{-\infty}^{\infty} \frac{dE}{2\pi} t_{\mathbf{rr}'} \text{Re}[\, G^<(\mathbf{r}, \mathbf{r}', E) \,] \qquad (7)$$

with $t_{\mathbf{rr'}}$ being the electronic hopping integral between sites $\mathbf{r}$ and $\mathbf{r}$', $g_e = 2$ reflecting the spin-degeneracy of the electrons, and $G^<(\mathbf{r}, \mathbf{r'}, E)$ being the lesser Green's function between sites $\mathbf{r}$ and $\mathbf{r}$'. In the absence of the leads, $G^<$ can be obtained from the retarded Green's function via

$$G^<(\mathbf{r}, \mathbf{r'}, E) = -2in_F(E)\text{Im}G^r(\mathbf{r}, \mathbf{r'}, E) \qquad (8)$$

where $n_F(E) = (e^{\beta E} + 1)^{-1}$ is the Fermi distribution function with $\beta = 1/k_B T$. To induce a current flowing through the network, one chooses two different chemical potentials, $\mu_{L,R} = \pm eV/2$, in the left (L) and right (R) leads, resulting in a voltage $V$ across the system. We can use $I_{\mathbf{rr'}}$ to compute the current between any two sites that are connected by electronic hopping, thus obtaining the spatial form of current flow in the network. The above form of Eq.(7) implies that $t_{\mathbf{rr'}}\text{Re}[\,G^<(\mathbf{r}, \mathbf{r'}, E)\,]$ is essentially an energy resolved probability for an electron to flow from site $\mathbf{r}$ to $\mathbf{r}$'. On the other hand, the definition of $G^<$ in Eq.(8) yields another important physical relevance of the lesser Green's function: considering the energy integral over the local lesser Greens function

$$\int_{-\infty}^{\infty} dE\, G^<(\mathbf{r}, \mathbf{r}, E) = -2i \int_{-\infty}^{\infty} dE\, n_F(E)\text{Im}G^r(\mathbf{r}, \mathbf{r}, E) = 2\pi i \int_{-\infty}^{\infty} dE\, n_F(E)\text{N}(\mathbf{r}, E)$$

$$= 2\pi i\, n_e(\mathbf{r})$$

where in the last equality we have used the definition for $n_e(\mathbf{r})$, the number of electrons at site $\mathbf{r}$, we find that the lesser Green's function is also a measure for the local electron density.

Eq.(7) implies that to investigate the form of charge transport through a network, it is necessary to calculate the lesser Green's function for a given spatial (and electronic) structure of a network and in the presence of defects or interactions such as an electron-phonon or electron-electron (Coulomb) interaction. This can be achieved by computing the lesser Green's function perturbatively, i.e., order by order in the hopping amplitude $-t$, in the defect scattering strength, or the electron-electron or electron-phonon interaction strength by using the Dyson equation, as described in more detail below. Since we are considering finite systems in real space, this is best done by defining Green's function matrices, $\hat{G}^{r,a,<}$ in real space whose (**rr'**) elements are given by $G^{r,a,<}(\mathbf{r}, \mathbf{r}', E)$. Rewriting the Hamiltonian in Eq.(1) in matrix form

$$H = \sum_\sigma \Psi_\sigma^\dagger \hat{H}_0 \Psi_\sigma + H_{lead} \quad , \tag{9}$$

where $\Psi_\sigma^\dagger, \Psi_\sigma$ are row and column spinors, containing the creation and annihilation operators of the system, respectively, the Dyson equations take the compact form

$$\hat{G}^<(E) = \hat{G}^r(E)\left[(\hat{g}^r(E))^{-1}\hat{g}^<(E)(\hat{g}^a(E))^{-1} + \hat{\Sigma}^<(E)\right]\hat{G}^a(E) \tag{10a}$$

$$\hat{G}^r(E) = \hat{g}^r(E) + \hat{g}^r(E)\left[\hat{H}_0 + \hat{\Sigma}^r(E)\right]\hat{G}^r(E) \quad . \tag{10b}$$

Here, $\hat{\Sigma}^{<,r}$ are the lesser and retarded self-energies that describe the effects arising from the scattering off impurities or defects, or from the electron-electron or electron-phonon interactions. Moreover, $\hat{g}^{<,r,a}$ are the lesser, retarded and advanced Green's function matrices of the network in the absence of any hopping element (i.e, for $\hat{H}_0 = 0$) or interactions. They are diagonal matrices with $\hat{g}^r$ containing the elements $g_0^r(E) = \left(E - E_0 - eV_g + i\delta\right)^{-1}$ for all sites in the network and $g_l^{r,a}(E) = \mp i\pi N_0$ for all sites in the leads that are connected to the network. In this form of $g_0^r$, we have included the effects of a gate voltage $V_g$ which uniformly shifts all energy levels, and thus allows one to consider charge transport through different energy eigenstates. When investigating the transport through networks of quantum dots or molecules, $g_0^r$ needs to be replaced by the appropriate Green's function describing a single dot or molecule. Moreover, the form of $g_l^{r,a}$ represents the wide-band limit of the leads possessing a constant density of states, $N_0$. $\hat{g}^<$ contains the elements $g_0^<(E) = -2in_F(E)\text{Im}g_0^r(E)$ and $g_l^<(E) =$

$-2in_F^{L,R}(E)\mathrm{Im}g_l^r(E)$, where $n_F^{L,R}(E) = n_F(E - \mu_{L,R})$ are the Fermi distribution functions in the left and right leads, respectively. It is the difference in the chemical potentials $\mu_{L,R}$ entering $g^<$ that gives rise to the flow of charge through the network.

To understand the origin of the Dyson equation, we consider the simple case of how the Green's function, $\hat{G}^r$, in a network in which electrons hop between sites is obtained from the Green's functions $\hat{g}^r$ in a network in which the hopping is absent. (here, we neglect any self-energy correction $\hat{\Sigma}$ in Eq.(10) that arise, for example, from the electron-phonon interaction). This is achieved through a perturbative inclusion of hopping processes which are described by the Hamiltonian matrix $\hat{H}_0$[cf. Eqs.(1) and (9)]. The resulting perturbative expansion of the full Green's function, $\hat{G}^r$, in form of Feynman diagrams is presented in Fig.2(c). The "cross" represents hopping processes between nearest neighbour sites **r** and **r'**, and the diagrams on the right hand side of the first equality represent the contributions to $\hat{G}^r$ to zero, first, second, etc. order in $\hat{H}_0$. This infinite series can be simplified by using the very definition of $\hat{G}^r$, yielding the second equality, which is the diagrammatic representation of the Dyson equation for $\hat{G}^r$ shown in Eq.(10b) (with $\hat{\Sigma} = 0$). In a similar fashion, one can incorporate the self-energy arising from the electron-phonon interaction, shown in Fig.2(d) (see discussion in Sec.2.2), perturbatively in the calculation of the full Green's function, yielding Eq.(10).

While the expression for $I_{\mathbf{rr'}}$ in Eq.(7) fully describes the charge transport through a network, it does not directly reveal some of its more salient features. However, important insight into the relation between the total current flowing through a network (which we will refer to as the network's global transport properties), the local flow of currents in the network (which we will refer to as the network's local transport properties) and the network's electronic structure, as reflected in the form of the above-defined Green's functions, can be gained by considering a non-interacting network with $\hat{\Sigma}^{<,r} = 0$. In this case, one can obtain a simplified expression from Eq.(7) for the total current flowing through a network, which is given by

$$I_c = 4g_e \frac{e}{\hbar} (\pi N_0)^2 t_l^4 \int_{-\infty}^{\infty} \frac{dE}{2\pi} |G^r(\mathbf{L}, \mathbf{R}, \omega)|^2 [n_F^L(E) - n_F^R(E)] \quad . \tag{11}$$

This expression reveals two important results. First, it implies that the total current through the systems depends only on the non-local retarded Green's function between the two sites **L** and **R** that are connected to the left and right leads, respectively [see Fig.2(a)]. In the limit of vanishing coupling to the leads, $t_l \to 0$, this Green's function can be computed using Eq.(5) with the wave-functions and energies given in Eqs.(3) and (4), respectively. Thus, one can calculate the global transport properties of the network from its equilibrium electronic structure. Second, in the limit $T \to 0$, the Fermi functions entering the integrand in Eq.(11) imply that only those states whose energies lie between $\mu_R < E < \mu_L$ [schematically shown as red lines in Fig.2(b)] contribute to the total current through the network. Since $\mu_{L,R} = \pm eV/2$, this also implies that when $V$ is increased and a new state enters the energy range between $\mu_L$ and $\mu_R$, $I$ increases sharply, giving rise to a step-like $IV$-curve, as discussed in more detail below.

The second important insight that can be gained from $I_{\mathbf{rr'}}$ pertains to the spatial nature of charge transport. Assuming again a non-interacting network, one obtains from Eq.(7) for the current flowing between sites **r** and **r'** in the network

$$I_{\mathbf{rr'}} = 2g_e \frac{e}{\hbar} t_{\mathbf{rr'}} \int_{-\infty}^{\infty} \frac{dE}{2\pi} t_l^2 (\pi N_0) [n_F^L(E) - n_F^R(E)]$$

$$\times \text{Im}[G^r(\mathbf{r}, \mathbf{L}, E) G^a(\mathbf{L}, \mathbf{r'} E) - G^r(\mathbf{r}, \mathbf{R}, E) G^a(\mathbf{R}, \mathbf{r'} E)] \quad , \tag{12}$$

where $-t$ is the electronic hopping amplitude between sites $\mathbf{r}$ and $\mathbf{r}'$. Since the Green's function $G^r(\mathbf{r}, \mathbf{L}, E)$ describes the propagation of an electron between sites $\mathbf{r}$ and $\mathbf{L}$, the above expression allows for an interesting spatial interpretation regarding the nature of the charge current. In particular, the form of the integrand implies that an electron taking part in the current between sites $\mathbf{r}$ and $\mathbf{r}'$, does not directly move between these two sites. Rather, the electron propagates from $\mathbf{r}$ and $\mathbf{r}'$ via the leads. Consider, for example, the contribution to the current arising from the part of the integrand given by

$$t_l^2(\pi N_0) \text{Im}[G^r(\mathbf{r}, \mathbf{L}, E) \, G^a(\mathbf{L}, \mathbf{r}'E)]$$

$$= -\text{Im}[G^r(\mathbf{r}, \mathbf{L}, E)(-t_l) \text{Im} g_l^r (-t_l) G^a(\mathbf{L}, \mathbf{r}'E)] \quad . \tag{13}$$

This contribution can be interpreted as an electron first propagating from site $\mathbf{r}$ to site $\mathbf{L}$ (the site the left lead is connected to), then hopping (via $-t_l$) onto the left lead with Green's function $g_l^r$, then hopping back to $\mathbf{L}$ (via $-t_l$), and finally propagating from site $\mathbf{L}$ to site $\mathbf{r}'$, as schematically shown in Fig. 3. The second term in the integrand in Eq.(12) describes the corresponding process involving the right lead. Interestingly enough, both processes lead to the same contribution to the current due to the identity

$$\text{Im}[G^r(\mathbf{r}, \mathbf{L}, E) \, G^a(\mathbf{L}, \mathbf{r}'E)] = -\text{Im}[G^r(\mathbf{r}, \mathbf{R}, E) \, G^a(\mathbf{R}, \mathbf{r}'E)] \quad .$$

Since the flow of charge between sites $\mathbf{r}$ and $\mathbf{r}'$ involves the propagation of electrons to the leads, and thus does not only involve the local properties of the network around sites $\mathbf{r}$ and $\mathbf{r}'$, but also those far removed from $\mathbf{r}$ and $\mathbf{r}'$, it is referred to as *non-local transport*. This type of transport is very different from the one in the classical regime, where the current between two sites only depends on local properties, namely, the voltage difference and local conductance between the sites.

## *2.2 Electron-Phonon Interaction and Dephasing*

As we discussed in the introduction, a transition from quantum mechanical to the classical transport behaviour can be characterized by the relation between the inelastic mean free path $l$ and the linear size $L$ of the system: for $l \gg L$, the transport is quantum mechanical, while for $l \ll L$ it is classical. A finite mean free path is created by any interaction that gives rise to inelastic scattering of the electrons. The simplest such interaction that allows a microscopic treatment is given by the electron-phonon interaction. We therefore consider the case where electrons interact at each site of the network with a local phonon modes (sometimes also referred to as a phonon bath), which is described by the Hamiltonian [39]

$$H_{e-ph} = g \sum_{\mathbf{r},\sigma} c_{\mathbf{r},\sigma}^\dagger c_{\mathbf{r},\sigma}(a_{\mathbf{r}}^\dagger + a_{\mathbf{r}}) + \omega_0 \sum_{\mathbf{r},\sigma} a_{\mathbf{r}}^\dagger a_{\mathbf{r}} \quad, \tag{14}$$

where $\omega_0$ is energy of the local phonon mode, $g$ is the electron phonon interaction strength, and $a_{\mathbf{r}}^\dagger, a_{\mathbf{r}}$ and are the bosonic operators that create or destroy a phonon at site $\mathbf{r}$. The general solution of such a model is quite difficult, due to the nature of the coupled Dyson-equations in Eqs.(10a) and (10b). However, they can be significantly simplified in the so-called high-temperature limit defined via $k_B T \gg \omega_0$ [39] (here, we set $\hbar = 1$). Here, only terms to leading order in $n_B(\omega_0) \approx k_B T/\omega_0$ are retained in the self-energies yielding

$$\hat{\Sigma}_{\mathbf{rr}}^{<,r}(E) = ig^2 \int_{-\infty}^{\infty} \frac{d\varepsilon}{2\pi} D^<(\varepsilon) \hat{G}_{\mathbf{rr}}^{<,r}(E - \varepsilon) \quad, \tag{15}$$

where

$$D^<(\varepsilon) = 2in_B(\varepsilon) \text{Im} D^r(\varepsilon) \tag{16a}$$

$$D^r(\varepsilon) = \frac{1}{\varepsilon - \omega_0 + i\delta} - \frac{1}{\varepsilon + \omega_0 + i\delta} \tag{16b}$$

are the lesser and retarded phonon Green's functions, and $n_B(\varepsilon) = 1/(e^{\beta\varepsilon} - 1)$ is the Bose distribution function. We assume that the phonon Green's functions remain unchanged in the presence of an applied bias [40]. Note that the self-energies are local, since phonon modes on different sites are not coupled. Inserting Eq.(16a) into Eq.(15) and taking the limits $\omega_0 \to 0$, and $k_B T \to 0$ with $k_B T/\omega_0 = const.$ yields

$$\hat{\Sigma}^{<,r}_{\mathbf{rr'}}(\omega) = 2g^2 \frac{k_B T}{\omega_0} \hat{G}^{<,r}_{\mathbf{rr'}}(\omega) \equiv \zeta \hat{G}^{<,r}_{\mathbf{rr'}}(\omega) \quad . \tag{17}$$

Thus the effective strength of the electron-phonon interaction can be described by a single parameter, $\zeta$. As such, we can now study the evolution of the network's electronic structure and transport properties from the ballistic, quantum mechanical limit to the diffusive, classical limit by increasing $\zeta$. As we show below, in the limit $\zeta \to \infty$, the network's transport properties map onto those of a classical resistor network.

The scattering (or dephasing) time that arises from the electron-phonon interaction is determined by $\zeta$, and can in general be extracted from the energy broadening of the electronic states: for simple networks and in the limit of small $\zeta$, one finds that the half width at half maximum of the energy broadened states is given by $\Gamma = \sqrt{p\zeta}$ where the dimensionless number p depends on the spatial structure and size of the network (for example, for a network with a single site, $p = 3$). Because of this non-universal behaviour, we define an effective dephasing rate $\Gamma/\hbar \equiv \sqrt{\zeta}/\hbar$ by setting $p = 1$ (at the most, this will introduce an error of order unity). The corresponding dephasing time (or lifetime) is then given by $\tau = \hbar/\Gamma$. Note that since the phonons give rise to point-like (i.e., isotropic) scattering, the transport and scattering times are identical.

After this brief review, we are now able to explore the nature of classical and quantum mechanical charge transport in more detail.

## 3. Classical Transport: Ohm's Law and IV curves

Ohm's law

$$I = \frac{1}{R}V = GV \qquad (18)$$

is the fundamental relation that allows us to connect the charge current $I$ flowing through a classical system with resistance $R$ or conductance $G$ to the voltage difference $V$ across the system. Since $G$ does not depend on $V$, Ohms law represents a linear relation between the current and the voltage, which is one of the characteristic hallmarks of classical transport. This allows us to obtain the total conductance of a network simply from the slope of the corresponding $IV$-curve, as shown in Fig. 4(a). As mentioned in the introduction, the microscopic origin of the resistance lies in the scattering of electrons off imperfections, defects or impurities in the lattice structure, or in the scattering arising from electron-phonon and electron-electron interactions. Within the Drude theory [41], the conductivity of the system is related to these scattering processes via the relation

$$\sigma = \frac{n_e e^2 \tau_{tr}}{m}, \qquad (19)$$

where $n_e$ is the electron density, $m$ is the electron mass, and $\tau_{tr}$ is the transport time. If the electronic scattering is isotropic in space (i.e., there is no preferred direction in which electrons are scattered) then the transport time and the lifetime of the electrons are identical. As we will see below, a comparison of the transport properties in the quantum and classical regimes will require us to compute how a classical current flows through a connected network of classical resistors. To this end, we need besides Ohm's law, a second set of rules known as Kirchhoff's laws [42]. Kirchhoff's first law states that the total current through a node of the network is zero (keeping in mind that the current is a directed quantity, with current flowing into a node having a positive sign, while the current flowing out of the node is negative). This rule simply reflects current conservation, requiring that the total current flowing into a node be equal to the total current flowing out of a node. The second of Kirchhoff's laws states that the directed sum of all potential differences around any closed loop in the network is zero.

A classical resistor network is in structure similar to the quantum network shown in Fig.2(a), albeit with the difference that the links the between nodes now represent classical resistors, which we assume to be equal for all links. Using a generalization of Kirchhoff's laws for arbitrary network geometries [42], we can now compute the spatial form of the current (i.e., the spatial current pattern) through such networks. This spatial form depends strongly on the relation between the system size and the size of the leads. If the leads are much narrower than the network [Fig.4(b)], the current exhibits a significant spatial variation in the network, while the charge flow is nearly uniform in a network attached to wide leads [Fig.4(c)]. Since the total current through the classical resistor network, as well as the currents through individual links (i.e., resistors), increases linearly with $V$, the overall spatial current pattern remains unaffected by changes in $V$. Finally, the current flowing through a link depends only on the bias difference between the adjacent nodes, and the resistance of the link, i.e., on local properties, and one therefore refers to charge transport in classical systems as local in nature. When a defect is inserted into such classical resistor network [Fig.4(d)], the changes in the current pattern only occur in the immediate vicinity of the defect, again reflecting the local nature of transport.

The linear form of $IV$-curves as well as the spatial form of the current patterns are two characteristic signatures of classical transport. As we will show in the following sections, they are drastically different from those of quantum networks.

## 4. Ballistic Quantum Mechanical Transport

In section 2, we gave a brief overview of the non-equilibrium Keldysh Green's function formalism which allows us to study the transport properties of quantum networks. In the following sections, we will identify the characteristic features of quantum transport using this formalism.

### *4.1 Charge Transport in Finite Quantum Networks*

To identify the most prominent features of quantum mechanical charge transport, we consider a non-interacting quantum network with $N_{x,y} = 31$ that is attached to two narrow leads, as shown in Fig.2(a). To gain insight into its electronic structure, we consider the local density of states (LDOS) $N(\mathbf{L}, E)$ at site $\mathbf{L}$ shown in Fig.5(a). We assume that the network contains one electron per site, implying that the chemical potential is located at $\mu = 0$, and that the LDOS is particle-hole symmetric. As expected for a network of finite size (and as discussed in Sec.2.1), the LDOS exhibits a series of sharp peaks, reflecting the presence of discrete energy states (the non-zero energy width of these states arises from the coupling to the leads). This discrete nature of energy levels in nanoscale systems, with a size-dependent energy spacing, has been experimentally observed in metal particles[43], quantum dots [34, 35, 44] and quantum point contacts [45]. In the right section of Fig.5(a), we have overlain a plot of the network's $IV$-curve, i.e., a plot of the total current through the network, $I_c$, as a function of the voltage $V$. We here make use of the fact that for a given $V$, all states that lie between $E = -eV/2$ and $E = +eV/2$ contribute to the current flowing through the system. Hence, when $V$ is increased, $I_c$ exhibits a sharp, nearly step-like increase every time $eV/2$ crosses a state with energy $E_\mathbf{k}$, as follows from a comparison of the LDOS and $I_c$ in Fig.5(a). The step-like form of $IV$-curves has been experimentally observed in metal particles [43, 46, 47], and is a direct signature of the discrete nature of the energy levels. It is qualitatively different from the linear form of $IV$-curves in the classical limit, and can thus be considered one of the distinguishing features of quantum transport. The small rounding of the steps in the $IV$-curve arises from the network's coupling to the leads and the resulting energy broadening of the states.

We next want to explore the spatial form of a current that is carried by a given energy state. To this end, we consider a sufficiently small voltage $V$, such that only a single energy state lies between $\pm eV/2$. The resulting current pattern for the $E_0 = 0$ state, shown in Fig.5(b), exhibits a spatially highly confined path that flows along a diagonal direction, and specularly reflects off the edges of the network. To understand the origin of this spatial form, we first note that the $E_0$ state possesses an $(N = N_{x,y})$-fold degeneracy, since there exist $N$ eigenstates with different quantum numbers $\mathbf{k} = (k_x, k_y)$ and the same energy $E_0$ [as follows from the analysis of $E_\mathbf{k}$ in Eq.(4)]. The corresponding wave-vectors $\mathbf{k}$ of these states are plotted in Fig.6(a), and form a diamond-like shape in the Brillouin zone. An electron with energy $E_0$ propagating through the network is composed of a superposition of these eigenstates, and the group velocity of its wave-packet is given by

$$\mathbf{v} = \frac{\partial E_\mathbf{k}}{\partial \mathbf{k}}, \qquad (20)$$

which points along the diagonal directions, thus explaining the direction of current flow through the network. However, this argument is based on a quasi-classical notion of an electron described as a wave-packet of sufficiently small size (thus almost resembling a classical particle), and thus does in general not explain the spatial current patterns associated with other energy states. Consider for example, the spatial form of the current carried by the state at $E_1 = 0.495t$, shown in Fig.5(c) (the latter was obtained by applying a gate voltage $V_g = E_1/e$ to the network). The spatial current pattern is qualitatively different from that of the $E_0 = 0$ state, with the current being distributed over much larger parts of the network. Interestingly enough, this spatial pattern reveals another quantum phenomenon – the backflow of currents [48], i.e., the flow of charge opposite to the applied voltage. This is particularly evident when one zooms into the center of the current pattern of Fig.5(c) [which is shown in Fig.6(b)], which reveals that while the overall charge current flows from left to right since $\mu_L > \mu_R$, the current in the center of the network flows in the opposite direction. The fact that the current carried by each energy state exhibits a qualitatively different spatial pattern, which in turn all differ from those of a classical resistor network [Fig.4(b) - (d)], and the existence of current backflow, represent additional distinguishing features of quantum mechanical transport. Finally, we note that the form of the spatial current patterns in Figs. 5(b) and (c) does not coincide with the spatial form of the local density of state, $N(\mathbf{r}, E)$ for the same energy states, shown in Figs. 5(d) and (e).

The question naturally arises to what extent the form of the coupling to the leads, and in particular the width of the leads, affects the results discussed above. The *IV*-curve of a quantum network connected to wide leads [Fig.7(a)] retains the same non-linear, step-like form, which thus remains a characteristic signature of quantum transport. However, the spatial current pattern is quite different from those shown above in that it reveals an almost uniform flow of charge across the system [Fig.7(b)]. This nearly uniform flow can be easily understood as arising from a superposition of currents patterns for the case of narrow leads. While the presence of narrow leads selects specific spatial channels for current flow, a superposition of all of these channels washes out the current patterns and leads to the near uniform flow of charge shown in Fig.7(b). Since this pattern is similar to that obtained in the classical limit, spatial current patterns in networks attached to wide leads are not necessarily a clear reflection of the quantum mechanical nature of charge transport.

*4.2 Disorder and Localization in Quantum Transport*

One of the most fundamental phenomena characterizing the quantum nature of transport in macroscopic systems are localization effects arising in the presence of disorder [11, 12]. Much progress has been made in understanding the nature of weak and (strong) Anderson localization effects, and it is beyond the scope of this article to review the large body of work that has contributed to this understanding (for some reviews, see Ref.[15-17]). We therefore just briefly review its most salient features. In the absence of any processes leading to electronic dephasing, the scattering of electrons by static disorder leads to an increase in the resistivity of the system for weak scattering (a phenomenon known as weak localization[12]), and to complete localization of the electronic wave-functions for strong disorder scattering, hence known as strong or Anderson localization [11]. These phenomena arise from the destructive interference between scattered electronic wave-functions on time-reversed paths. As a result, in the case of Anderson localization, the non-local Green's function which determines the total current through the system [Eq.(11)] decays exponentially with distance between the attached leads, leading to an exponentially suppressed conductance of the system. The characteristic length scale of this exponential decay is known as the localization length.

In a finite network, the effects of impurities on the system's transport properties are quite complex since they sensitively depend on the interplay between the form of the spatial current pattern and the location of a defect. To demonstrate this, let us again consider the spatially highly-confined current pattern carried by $E_0 = 0$ state, shown in

Fig.8(a). Inspired by our classical notion of current flow, one might wonder whether a defect (i.e., an obstacle) that is placed in the path of the current (e.g., at the position indicated by the filled blue circle) might have a different, and presumably stronger effect on the network's transport properties than a defect placed in the middle of the network (e.g., at the positions indicated by the open yellow circles), where the current density is essentially zero. Indeed, we find that placing defects at the locations of the open yellow circles does neither change the total current through the network nor the spatial current pattern of the $E_0$ state. While this result might be intuitive on the classical level, its understanding requires some deeper insight into quantum mechanics. On the quantum level, this result is actually quite puzzling since the LDOS of the $E_0$ state [see Fig.5(d)] is non-zero at the sites $\mathbf{r}_0$ indicated by the open yellow circles, implying that electrons in this state should be scattered by the defect. However, to evaluate the effect of defect scattering on the charge transport through the $E_0$ state, one needs to consider the quantum mechanical amplitude for an electron to propagate from site $\mathbf{L}$ to $\mathbf{r}_0$, which is given by the non-local Green's function $G^r(\mathbf{L}, \mathbf{r}_0, E)$. It turns out that this amplitude vanishes due to a destructive interference between the different states $|\psi_\mathbf{k}\rangle$ that are located at $E_0 = 0$, such that the defect at $\mathbf{r}_0$ does not affect the charge current entering the system at $\mathbf{L}$, and thus the transport properties of this state.

In contrast, when a defect is placed into the path of the current at $\mathbf{r}_1$ [indicated by the filled blue circle in Fig.8(a)], the effect on the system's transport properties are even more striking than anticipated: the conductance rather than decreasing by a factor of two because only one of the two current branches is blocked by the defect, actually decreases by more than 5 orders of magnitude. A very small residual conductance remains due to the network's coupling to the leads (which induces some residual dephasing, see Sec. 5.1). Insight into the origin of this dramatic change is gained by considering the effect of the defect on the overall electronic structure of the network. Indeed, a comparison of the LDOS $N(\mathbf{L}, E)$, for the networks with and without a defect shown in Fig.8(b) shows that the $E_0$ state has been shifted away from zero energy in the network containing a defect. In contrast to the case where the defect was located at $\mathbf{r}_0$, the amplitude for electrons to propagate from $\mathbf{L}$ to the site of the defect at $\mathbf{r}_1$ does not vanish. The electrons in the $E_0$ state are therefore scattered by the defect, resulting in destructive interference which shifts the energy of this state away from the Fermi energy. Since the chemical potentials in the left and right leads remain unchanged, no state of the system lies any longer between $\mu_L$ and $\mu_R$, leading to a dramatic decrease in the network's conductance.

While the above results are consistent with our classical notion of currents as the flow of particle-like electrons, quantum mechanics always throws a "curve-ball" upsetting our classical understanding. Consider for example, the spatial current pattern carried by the state at $E_1 = 0.435t$ whose spatial current pattern is shown in Fig.8(c). This spatial pattern also reveals a large current density at $\mathbf{r}_1$, such that when a defect is placed at this site, it is not unexpected that a significant change in the spatial current pattern occurs, as shown in Fig. 8(d). However, the total current carried by this state is entirely unaffected, and remains the same as that in the network without a defect. This observation is quite perplexing on a classical level, but allowed by quantum mechanics, since the total current through the system and the form of the spatial current patterns are determined by different combinations of non-local Greens functions [cf. Eqs.(11) and (12)]. While the Green's functions involving the total current through the system, Eq.(11), remain unchanged by the presence of the defect (due to the vanishing of the non-local Green's function between $\mathbf{L}$ and $\mathbf{r}_1$), the Green's functions determining the spatial current pattern $I_{\mathbf{rr'}}$ in Eq.(12) are altered. The observation that the charge transport of certain states is affected by the presence of a defect, while that of other states is not, is of fundamental importance since it implies that defects can be *gated-away*: by gating the network and hence selecting different states for current transport, it is possible to make defects *invisible* to the global current transport through the system.

## 5. Crossover from Quantum Mechanical to Classical Transport

In the preceding sections, we had shown that the transport properties of networks in the quantum and classical limits are qualitatively different, with properties such as the $IV$-curves or the spatial current patterns exhibiting significant differences, or phenomena such as Anderson localization existing only in the quantum limit. We next discuss how the crossover between these two opposite limits of transport behaviour can be theoretically described, and in doing so, identify its signatures. This crossover from ballistic to diffusive transport occurs when the inelastic mean-free path evolves from being much larger than the network size, to being much smaller. To model this transition and to introduce a finite mean free path in the system, we consider the scattering of electrons by local phonon modes (see section 2.2) which is described by the Hamiltonian

$$H_{e-ph} = g \sum_{\mathbf{r},\sigma} c^\dagger_{\mathbf{r},\sigma} c_{\mathbf{r},\sigma}(a^\dagger_\mathbf{r} + a_\mathbf{r}) + \omega_0 \sum_{\mathbf{r},\sigma} a^\dagger_\mathbf{r} a_\mathbf{r} \quad . \tag{21}$$

The crossover behaviour in the transport properties can then be explored as a function of the dephasing rate $\Gamma/\hbar = \sqrt{\zeta}/\hbar$ where $\zeta = 2g^2 k_B T/\hbar\omega_0$ represents an effective electron-phonon scattering strength. By increasing $\Gamma$, the mean free path is reduced, allowing us to study the evolution from $l \gg L$ to $l \ll L$.

*5.1 Crossover with increasing dephasing rate*

We begin our study of the crossover between quantum and classical transport by considering the effects of increasing electronic dephasing on the network's non-linear $IV$-curve, as shown in Fig.9(a). The evolution of the $IV$-curve with increasing $\Gamma$ exhibits two interesting features. First, the step-like form of the $IV$-curve for $\Gamma=0$ is smoothed out with increasing $\Gamma$, such that the current eventually increases linearly with the applied voltage, and recovers the linear $IV$-curve characteristic of a classical system. Since the step-like form of the $IV$-curves is a direct consequence of the discrete nature of the electronic states in the quantum limit, these results suggest that the LDOS evolves to a more uniform structure in energy. This conclusion is borne out by the form of the LDOS, $N(\mathbf{L}, E)$ presented in Fig.9(b). Here, we see that with increasing $\Gamma$, the peaks in the LDOS are broadened in energy, and begin to overlap, until already for $\Gamma = 0.3t$, the density of states is nearly uniform in energy, reflecting the nearly linear form of the $IV$-curve shown in Fig.9(a).

The second interesting feature is the decrease in the slope of the $IV$-curve, representing the overall conductance $G$ of the network, with increasing $\Gamma$. This reduction in $G$ is expected since increased scattering of electrons by phonons randomizes the electronic trajectories, hence reduces the drift velocity of the electrons, and thus increases the resistance of the network. A plot of $G$ as a function of the dephasing (or lifetime) $\tau = \hbar/\Gamma$ [Fig. 9(c)] reveals that for small $\tau$ (i.e., in the classical limit), $G \sim \tau$, as expected from the Drude model.

Complementary insight into the evolution of the transport properties is given by the transformation of the spatial current patterns with increasing $\Gamma$, as shown in Fig.10(a)-(f) for the current pattern of the $E_0$-state. For small $\Gamma$ [Fig. 10(a)], when the mean free path is still much larger than the system size, the spatial current pattern is similar to the one shown in Fig.5(b), with the current following an almost ballistic path. For larger $\Gamma$, however, the current pattern becomes increasingly more diffusive with increasing distance from the leads [see yellow arrow in Fig 10(c)]. This change in current pattern occurs when the mean-free path $l$ becomes smaller than the system size and the electrons are scattered at least once while traversing the network, thus signifying the crossover to classical transport. Since scattering destroys the electronic phase coherence, and thus randomizes the electron trajectory, the current pattern becomes more diffusive after a distance of $l$ from the leads. Note, that for intermediate values of $\Gamma$ [Fig.10(b),(c)], the network exhibits classical global transport properties – a linear $IV$-curve -- but local quantum properties. Increasing $\Gamma$ further reduces the mean-free path, and thus the size of the region around the leads where a coherent current pattern can be found. Concomitant with the transformation to a more diffusive current pattern, there is a redistribution of current density towards the center of the network, as shown in Figs. 10(c) - (f), which becomes the prominent path for sufficiently large $\Gamma$ [Fig. 10(f)]. This redistribution of the current density is particularly evident when one plots the current flowing along horizontal links in the center column of the network [Fig.10(g)] which is indicated by a dashed red line in Fig. 10(e). Concomitant with the change in the current pattern, the current carried by the $E_0 = 0$ state decreases rapidly with increasing $\Gamma$, as shown in Fig. 10(h), again reflecting the increased resistance of the network.

To determine whether the spatial current pattern, similar to the $IV$-curve, reflects the crossover to classical transport behaviour, we compare the spatial current pattern of the $E_0$ state at large $\Gamma = 10t$ [Fig.11(a)] with that of a classical resistor network [Fig.11(b)]. For the calculation of the latter [42], we have assumed that all resistances between neighbouring nodes in the network are equal. These two current patterns are all but indistinguishable, as can be seen by comparing the currents flowing along horizontal links in the middle column, as shown in Fig.11(c). Interestingly enough, one obtains overall good agreement with the classical current pattern (with the exception of the network's edges) already for $\Gamma = 2t$. In the same limit, we also find that the normalized current pattern, i.e., $I_{\mathbf{rr'}}/I_c$ becomes independent of $\Gamma$, as expected for a classical resistor network.

Finally, we showed above that the spatial current patterns in the quantum limit vary considerably between different energy states. In contrast, in the classical limit, there exists only the single current pattern of a classical resistor network. This suggests that all of the different spatial current patterns of the quantum limit evolve into the same classical current pattern. This evolution, however, reveals some surprising features, as demonstrated in in Fig 12, where we present the transformation of the spatial current patterns carried by the $E_1 = 0.495t$ state [for $\Gamma = 0$, the current pattern is shown in Fig.5(c)]. In particular, we observe that this current pattern does not directly evolve into that of the classical resistor network, but rather first evolves into the pattern of the $E_0 = 0$ state [Fig. 10], and then follows its transformation to that of the classical resistor network. This evolution is a direct consequence of an important physical effect of the electron-phonon interaction: by coupling different energy states, this interaction mixes their properties. Since the current flowing through the $E_0$ state is much larger (due to its degeneracy) than that of the $E_1$ state, the current through the latter eventually acquires properties of the current flowing through the $E_0$ state. Also note that concomitant with the evolution of the current pattern, the backflow of currents [see Fig.6(b)] vanishes.

The above results demonstrate that while we started with a purely quantum mechanical theory, the coupling of the electronic degrees of freedom to phonon modes smoothly transforms the network's transport properties into those of a classical system.

*5.2 Destruction of Anderson Localization by Dephasing*

Anderson localization in a disordered system arises from a destructive interference of scattered wave-functions between time-reversed paths. However, a destructive interference can only be achieved if the electrons remain phase coherent over the length of the backscattered paths. This implies that any interaction that destroys the electronic phase coherence will reduce the destructive interference between the scattered waves, thus suppresses localization and therefore increases the conductance of the system. In macroscopic system, the effects of various dephasing mechanisms on disorder-induced weak or Anderson localization have been considered [49-51] with electron-phonon coupling giving rise to variable-range hopping [51]. That dephasing actually increases the conductance of a disordered network is strikingly different from the behaviour found in clean networks, where the conductance *decreases* with increasing dephasing, as shown in Fig.10(h).

To demonstrate this rather perplexing behaviour, we return to the case considered in Fig.8 where a single defect at $\mathbf{r}_1$ was placed into the path of the current carried by the $E_0 = 0$ state. In Fig.13(a) we show how the total charge current, $I_c$, through the network changes with increasing $\Gamma$. For small $\Gamma$, the current increases, indicating the suppression of destructive interference, while for large $\Gamma$, the current decreases, similar to the case in a clean network. As a result, $I_c$, exhibits a maximum for some intermediate values of $\Gamma$ (this maximum will be of interest when discussing energy transport in biological systems, see Sec.7). It is interesting to note that the currents both for the clean system as well as the one containing the defect are essentially equal for $\Gamma > 0.5t$, implying that dephasing induced by phonons has essentially reversed the effect of the defect on the charge transport. This is also reflected in the form of the density of states: while for small $\Gamma$ [Fig.13(b)] $N(\mathbf{L}, E)$ in the vicinity of the

Fermi energy still differs significantly between the clean and the disordered network, $N(\mathbf{L}, E)$ in both networks is quite similar at large $\Gamma$ [Fig.13(c)].

The qualitatively different transport behaviour of the two systems for $\Gamma \ll t$, and their almost identical conductance for $\Gamma > 0.5t$ is also reflected in the form of the spatial current paths. In the former case, the spatial current patterns in the two systems differ significantly [cf. Figs. 14(a) and (b) for $\Gamma = 0.1t$] with the defect nearly completely blocking the upper current path. In the latter case, however, [cf. Figs. 14(c) and (d) for $\Gamma = 0.5t$], the spatial current patterns are all but indistinguishable. In this case, the impurity affects the current pattern only in its immediate vicinity, similar to the effects of defects in classical resistor networks [Fig. 4(d)]. This qualitatively different behaviour with the defect dramatically changing the network's global transport properties for $\Gamma \ll t$, but only affecting the local current pattern for $\Gamma > 0.5t$, reflects the qualitatively different nature of charge transport – non-local versus local – in the quantum and classical transport regimes.

*5.3 Crossover with increasing system size*

While we considered above the crossover between quantum and classical transport behaviour by increasing the strength of the electron-phonon interaction, thus decreasing both the dephasing time and the mean-free path, we can also investigate the crossover behaviour by changing the size of the system while keeping the mean free path constant. To demonstrate this, we consider in Fig.15(a) the evolution of the $IV$-curve with increasing dephasing rate $\Gamma$ for a smaller network with $N_{x,y} = 11$. As expected, we find that the $IV$-curve retains its step-like, non-linear form up to much larger values of $\Gamma$. In particular, for values of $\Gamma$ where the $IV$-curve of the larger network already exhibits a linear relationship (implying $l < L$), the one for the smaller network still possesses a non-linear form (and thus $l > L$). This behaviour is also mirrored in the spatial current pattern: while the spatial current pattern for the larger network already exhibits significant dephasing effects for $\Gamma = 0.6t$, the smaller network still exhibits a well-defined current pattern throughout the entire network [Fig15(b)]. These results demonstrate that the crossover between the quantum and classical regimes can occur both with increasing strength of the electron-phonon interaction (which can be driven by temperature effects), as well as through changes in the system size.

*5.4 Crossover in graphene, topological insulators or arrays of quantum dots*

The same crossover behaviour in transport properties discussed above can also be observed in materials that possess more complicated electronic bandstructures, such as graphene [30] or topological insulators [32]. Graphene is of particular interest since the long mean free path in this material allows to directly observe the quantum limit of transport behaviour. Interestingly enough, in a topological insulator, an increase in the dephasing rate does not only affect the *IV* curve or current patterns due to the randomization of the electron trajectories, but also leads to a destruction of its topological nature due to the loss of phase coherence [32]. Recent experimental advances in building networks of quantum dot with varying strength of the electronic hopping between the dots [52], provide a new path towards the experimental study of the crossover from quantum to classical transport.

*5.5 Quantum to classical crossover in the shot noise*

In our discussion of the crossover from quantum to classical transport, we focused on its signatures in the *IV*-curves and spatial current patterns. Another physical property related to charge transport in which this crossover has been extensively studied is the shot noise [18]. It was first investigated by Schottky as fluctuations of the current in vacuum tubes and arises from the discrete nature of the electronic charge. The investigation of shot noise in mesoscopic conductors has attracted significant interest [21, 53, 54]. It is a quantum phenomenon, arising from the wave-line nature of the electron and the resulting probabilistic transmission of an electron (as part of a current) through a system. As such, shot-noise vanishes in the classical limit [19]. Its evolution from the quantum to the classical limit has been theoretically predicted [20, 55] to occur in chaotic mesoscopic systems (such as cavities) when the dwell time of an electron in the cavity decreases. This prediction was subsequently confirmed experimentally [56]. Finally, we mention in passing that the crossover phenomena in transport have also been studied in the temperature dependence of the resistivity [57], as well as in the context of molecular electronics [58].

## 6. Spatial Imaging of Current Patterns: a view into the local transport properties of quantum and classical networks

In the preceding sections, we had shown that the crossover from quantum to classical transport behaviour is reflected in the spatial form of the current patterns. Thus gaining experimental access to these current patterns, and being able to image them in real space, would provide unprecedented insight into the transport properties of a system.

In mesoscopic systems, such imaging of current patterns has been achieved using a scanning probe microscope (SPM) [59-62]. In these experiments, the tip of the SPM is electrostatically charged and brought into close proximity to a system through which a current flows. The tip's electrostatic potential then scatters the electrons moving through the system, similar to the effect of defects discussed in Sec.4.2. In particular, when the tip is positioned above a region of high current density (such as the defect in Fig.8(a) was positioned directly in the path of the current), the total current through the system is strongly suppressed, while the current is only weakly affected when the tip is located over a region of low current density. Plotting the variations of the current through the system as a function of tip position then images the spatial regions of high and low current density, as shown in Fig.16(a). This experimental technique has proven to be invaluable to map out the spatial current patterns in mesoscopic systems, even succeeding in visualizing universal conductance fluctuations [63] and weak localization effects [64]. However, due to the size of the transverse width of the SPM's perturbing electrostatic potential, the spatial resolution of this technique is insufficient to detect the atomic length scale variations in the spatial current patterns shown above and predicted to exist in nanoscale systems [22, 24, 30, 65-68].

We therefore recently proposed a novel technique using a scanning tunnelling microscope (STM) [10]. By measuring the current flowing from the STM tip into a narrow lead attached to a nanoscale network as a function of tip position, we demonstrated that it is possible to image current patterns with a spatial resolution set by the lattice constant [see Fig. 16(b)]. Since this technique is also applicable in the presence of electronic dephasing, it is well suited to investigate the crossover from quantum to classical transport.

## 7. Crossover from Quantum to Classical Energy Transport in Biological Systems

### 7.1 Biological complexes as a Network Model

Biological processes occur in open systems where the interaction with many different degrees of freedom should inherently destroy the system's quantum nature through dephasing, and thus lead to classical behaviour. However, over the last decade, a series of experiments investigating the nature of excitonic energy transport in light-harvesting biological complexes have provided much evidence that this transport is neither fully classical nor fully quantum mechanical [9, 69]. Of particular interest has been the Fenna-Matthews-Olson complex (FMO) [70], a photosynthetic light-harvesting complex in green sulfur bacteria, which has served as a canonical model system to explore the nature of excitonic energy transfer. FMO transfers excitons -- particle-hole pairs – from the chlorosome, where they are created via absorption of photons, to the reaction center, where charge separation takes place and energy is stored [71]. This transport process is highly efficient [72] with the mechanism underlying this high efficiency remaining a topic of debate. Ultrafast spectroscopy experiments have reported that excitons created by short laser pulses exhibit quantum coherence for timescales much longer than expected for a classical system [9]. This has raised the intriguing possibility that Nature is employing quantum mechanics to increase the efficiency of transport processes.

To understand the nature of exciton transport in more detail, let us consider the excitonic structure of FMO. After absorption of a photon, an exciton is created in the chlorosome which enters the FMO and is transferred to the reaction center via the FMO's bacteriochlorophyll chromophores (BChls) [see Fig. 17(c)]. The exciton is believed to enter the FMO complex through BChls 1 and 6, the chromophores closest to the chlorosome and exit the FMO to the reaction center through BChl 3. Theoretical models of FMO and other photosynthetic protein-pigment complexes have suggested that the high efficiency of transport process arises from a coupling to the FMO's environment, which results in an interaction between the exciton and phonon modes residing in the environment [72-75]. This has given rise to the notion of

*environmentally assisted quantum transport* (EnAQT) [74, 75]. The similarity of this transport process to the electronic processes discussed above suggests that the development of a similar theoretical framework mapping the FMO complex onto a network model and using the Keldysh formalism to explore the transport of excitons through it, is appropriate [33]. The main difference to the case of charge transport is that an exciton can be annihilated in the network through recombination and emission of a photon.

We therefore model the flow of excitons through FMO by mapping the FMO onto a network of seven sites representing the seven chromophores of FMO. The chlorosome and the reaction center are represented by two leads through which excitons can enter and exit the FMO. While the supply of excitons from the chlorosome varies based on the environmental light levels, it is consistently low enough, such that at any time, the FMO contains at the most one exciton, such that exciton-exciton interactions can be neglected. At each chromophore site, there exists a non-zero probability for the exciton to either recombine (under emission of a photon) or to be scattered by a phonon. The first process leads to loss of excitons from the FMO, while the second one leads to dephasing of the excitons. The FMO complex is then represented by the Hamiltonian

$$H = H_{FMO} + H_{rec} + H_T + H_{ex-ph} + H_c + H_r \quad (22)$$

Here

$$H_{FMO} = \sum_{i=1}^{7} E_i d_i^\dagger d_i + \sum_{i \neq j} V_{ij} d_i^\dagger d_j \quad (23)$$

describes the excitonic structure of the FMO, with $E_i$ representing the excitonic on-site energies of the seven chromophore sites, $V_{ij}$ being the excitonic hopping amplitude between them, and $d_i^\dagger, d_i$ are the bosonic operators which create or annihilate an exciton at site $i$. The values for $E_i$ and $V_{ij}$ were determined using electron spectroscopy [76]. Moreover,

$$H_{rec} = \sum_{i=1}^{7} V_b \left( e_i^\dagger d_i + h.c. \right) \qquad (24)$$

describes the process of exciton recombination accompanied by the emission of a photon with $e_i^\dagger$ being the photon creation operator at site $i$. The coupling of the FMO complex to the chlorosome and reaction center, as described by $H_T$, as well as the exciton-phonon interaction, $H_{ex-ph}$, are similar to the respective interactions in the electronic case in Eqs.(1) and (14). Finally, $H_c$ and $H_r$ describe the excitonic structure of the chlorosome and the reaction center, which we assume to possess an incoherent spectroscopic structure represented by a constant spectral density. The transport of excitons through the FMO complex can now be investigated using the non-equilibrium Keldysh Green's function formalism [33], where the flow of excitons is driven by a difference in the exciton occupation number between the chlorosome and reaction center. This formalism allows us to seamlessly consider the evolution of the FMO's transport properties from the coherent quantum mechanical regime to the classical diffusive regime.

*7.2 Excitonic Energy Transport and the Noisy Environment*

Since the excitonic on-site energies as well as the coupling between excitonic sites in the FMO are highly non-uniform (i.e., disordered), we expect that the excitonic current flowing through the FMO exhibits a non-monotonic dependence on the dephasing rate similar to the behaviour of a charge current in a disordered network discussed in

Sec.5.2. This expectation is borne out by the explicit form of the computed exciton current, $I_{out}$, from FMO to the reaction center, which is shown in Fig.17(a). For small Γ, the dephasing suppresses the destructive interference of the disorder-scattered wavefunctions (which is the precursor of localization in macroscopic systems), thus leading to an increase in the excitonic current (or in other words, of the excitonic conductance). In contrast, in the limit of large dephasing rates, the scattering of excitons by phonons inhibits the flow of excitons, thus increasing the FMO's resistance (similar to the behaviour observed in charge transport), a phenomenon also attributed to the quantum Zeno effect [74]. As a result of these competing effects arising from scattering by phonons -- suppression of localization versus inhibiting the flow of excitons – the excitonic current through FMO exhibits a maximum near $\Gamma = 350 cm^{-1}$. The fact that the excitonic current can be increased by coupling to a noisy environment (i.e., the phonon bath) is the foundation for the notion of *environmentally assisted quantum transport* [73-75]. However, it is the same microscopic mechanism that underlies the behaviour of the charge current in disordered networks [Fig.13(a)], or the emergence of variable-range hopping in macroscopic systems. Since in contrast to charge currents, the excitonic current is not conserved due to the possibility for excitonic recombination, the excitonic current entering the FMO, $I_{in}$, is in general larger than $I_{out}$, as shown in Fig.17(b). The ratio $I_{out}/I_{in}$ can be defined as the efficiency for excitonic transport through the FMO.

    This crossover behaviour in the FMO's transport behaviour between small and large values of dephasing is also reflected in the spatial pattern of the exciton current between the chromophore sites, as shown in Fig.17(c), and the recombination rates shown in Fig.18. In the absence of dephasing ($\Gamma = 0$), which represents the coherent quantum mechanical limit for excitonic transport, the disorder in the on-site energies

and hopping amplitudes leads to weakly coupled excitonic sites. This results in weak overlap between the excitonic states at different sites, which inhibits transport and leads to trapping of excitons, in particular at sites 2, 5 and 6. This trapping implies that the likelihood for recombination is increased, leading to larger recombination rates at sites 2, 5 and 6, as shown in Fig.18. As a result, the number of excitons leaving the FMO complex to the reaction center is significantly reduced from that entering the FMO. With increasing dephasing, the overlap between states at different sites increases (due to the induced energy broadening of the states), and the excitonic transport through the FMO is enhanced. At the same time, excitons are less likely to be trapped in the FMO and thus to recombine, leading to a reduction in the recombination rate (see Fig.18). For large dephasing rates, corresponding to incoherent (i.e., classical) transport, the excitons flow along the paths where the inter-chromophoric couplings, $V_{ij}$, are largest [77, 78]. The subtle evolution of the spatial flow patterns of the excitons for intermediate dephasing rates reflects the transition from coherent to incoherent transport. Our findings demonstrate that transport through FMO depends strongly on the dephasing rate arising from a coupling to a noisy environment. Since the same dephasing rate also determines the lifetime of laser-induced coherence in FMO as observed in spectroscopic experiments.[9, 69, 79, 80], it represents the fundamental link between coherent transport and long-lived coherence in FMO. It is interesting to note that the inclusion of spatial correlations between phonon modes (which has not been considered here) leads to a reduction in the effective dephasing rate of the system [81, 82]. To what extent the above-described phenomena evolve in networks of light-harvesting complexes is also a question of great current interest [83, 84]. Finally, we would like to point out that while the energy scales in biological light-harvesting complexes and (disordered) nanoscopic or mesoscopic electronic systems can be very different, the qualitative effect of phonon-induced dephasing increasing the system's conductance is the same in both, as follows from a comparison of Figs.13(a) and 17(a).

## 8. Conclusions

Understanding the crossover from quantum to classical transport is not only of fundamental scientific interest, but also of great technological importance. Recent advances in creating transistors on the sub-10nm scale (with some being as small as 1nm), have made it clear that we have reached not only the end of the "classical rope" but also of Moore's law. This renders the investigation and understanding of the quantum-classical crossover a somewhat urgent matter: without it, we do not know whether the emergence of quantum phenomena (ranging from tunnelling to interference) will still allow us to perform classical logic operations in nanoscale transistors. On the other hand, the crossover regime, where global classical transport properties coexist with local quantum properties, might hold new opportunities for the emergence of fascinating physical phenomena and technological applications. Clearly, this is a topic deserving our attention.

The results discussed above have also raised an important question pertaining to Nature's exploitation of quantum mechanics in increasing the efficiency of biological processes. While it was shown that the excitonic current through FMO exhibits a non-monotonic dependence on the coupling to its environment, it is not clear whether Nature has evolved the "parameters" of the actual FMO complex and its environment such that the energy current is indeed close to the theoretically possible maximum. That it did might be reasonable to assume since the exciton current to the reaction center is likely relevant to evolutionary fitness in light-starved green sulfur bacteria [85]. From a physicist's point of view, this would correspond to the solution of an optimization problem: in this case, the conductance of the FMO is larger than it would be in the classical limit, allowing excitons to move more quickly through the FMO (which is important to avoid recombination and thus loss of the energy). On the other hand, for this optimal set of parameters, the FMO complex is protected from localization effects, arising from disorder which is intrinsic to biological systems. Clearly, the physics of the crossover regime between quantum and classical transport has much to offer both for technological applications as well as for understanding the evolutionary fitness of biological complexes.

**Acknowledgements:** DKM gratefully acknowledges support from the U. S. Department of Energy, Office of Science, Basic Energy Sciences, under Award No. DE-FG02-05ER46225. DKM also acknowledges stimulating discussions with T. Can, G. Engel, D. Goldhaber-Gordon, H. Jaeger, H. Manoharan, P. Guyot-Sionnest, and D. Talapin.

# Figures and Legends

# Figure 1

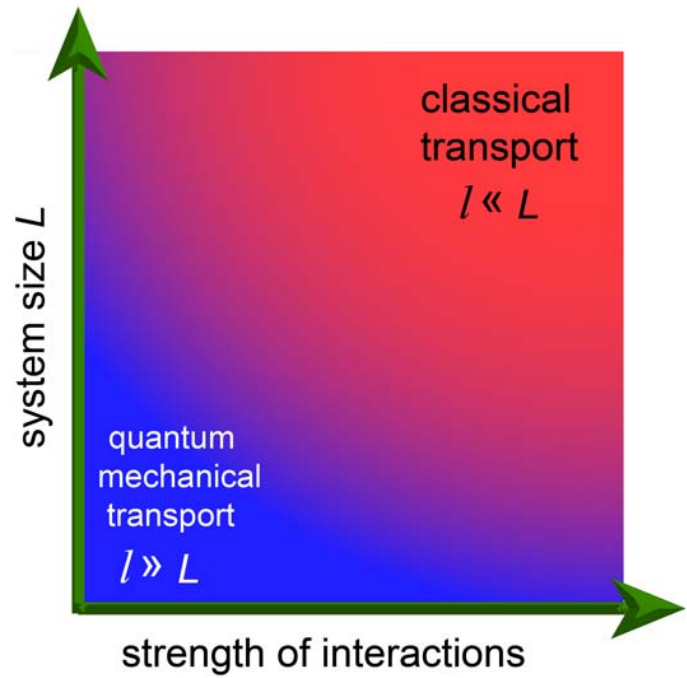

The crossover from quantum transport where the mean free path $l$ is much larger than the system size $L$, to classical transport where $l \ll L$, occurs with increasing system size or strength of interactions.

**Figure 2**

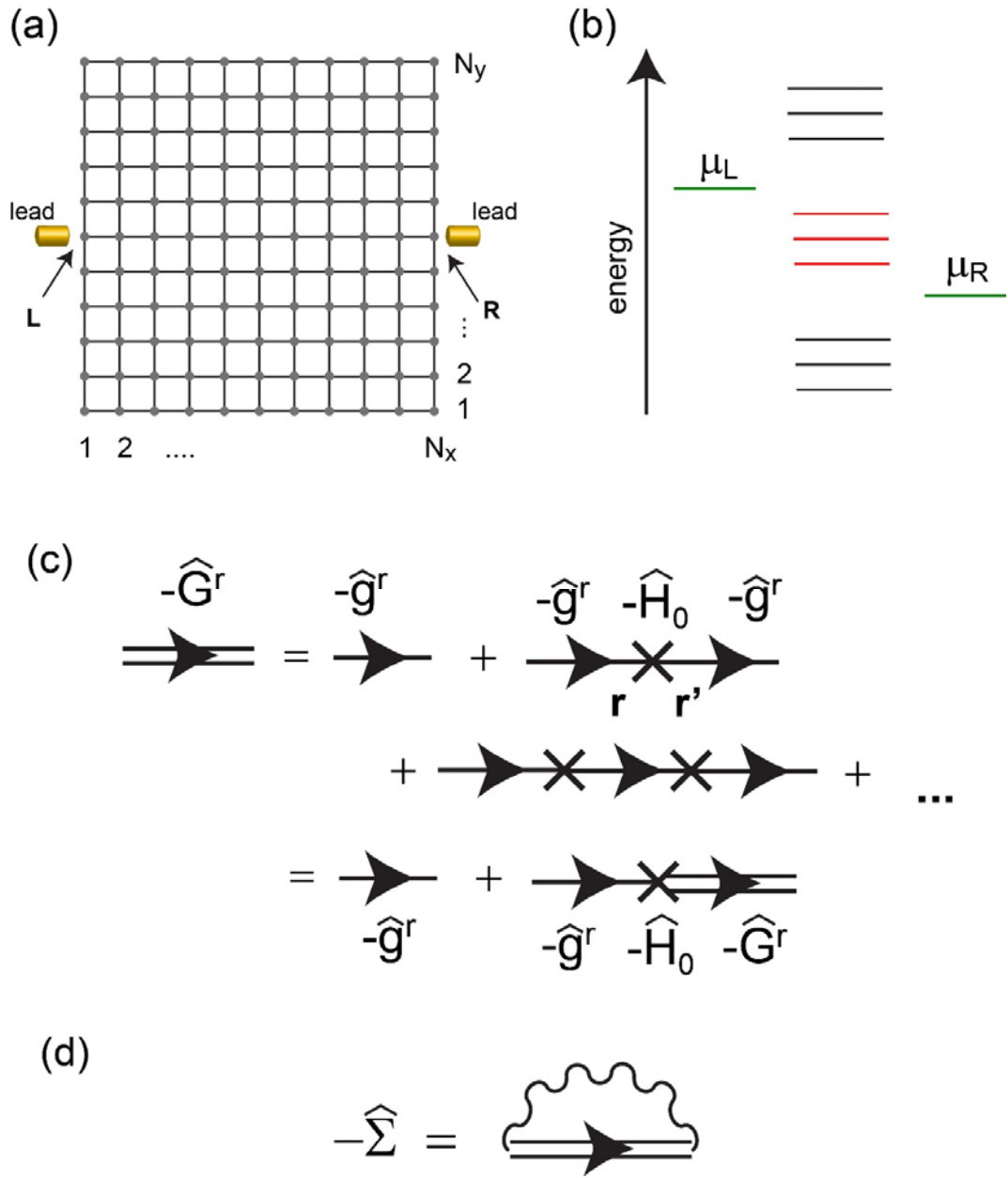

*(a)* Schematic representation of a network that is attached to narrow leads. The network sites attached to the leads are denoted by **L** and **R**. The gray dots are the nodes (sites) of the network representing atoms, molecules or quantum dots, and the lines representing the electronic hopping.

*(b)* Only those states (indicated by red lines) whose energies lie between the left and right chemical potentials $\mu_{L,R}$ take part in charge transport.

(c) Representation of the Dyson equation for $\hat{G}^r$ using Feynman diagrams. The diagram with the double lines represents the full Green's function $-\hat{G}^r$, while the diagrams with a single line represent $-\hat{g}^r$. The cross represents $-\hat{H}_0$ which describes the electronic hopping between sites $\mathbf{r}$ and $\mathbf{r}'$.

(d) Representation of the (self-consistent) second order self-energy correction, $\hat{\Sigma}^r$, arising from an electron-phonon interaction. The wavy line represents the phonon Green's function.

**Figure 3**

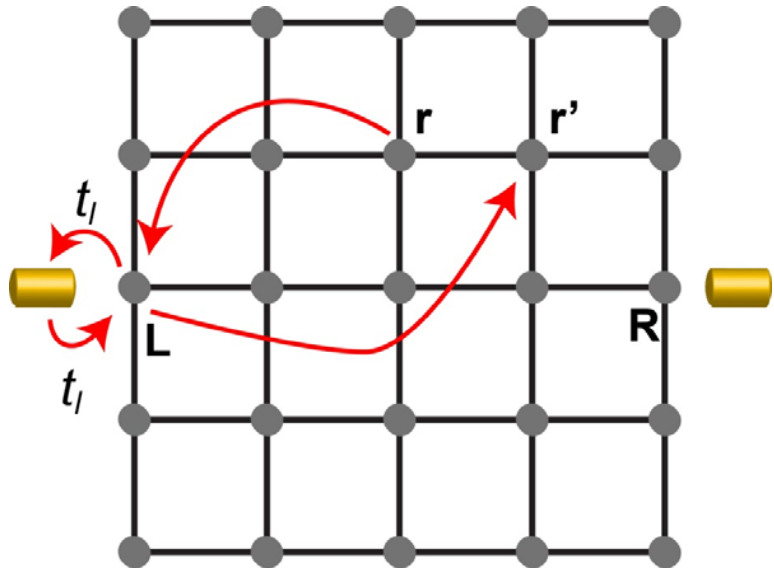

Schematic representation of the path of an electron contributing to the current between sites **r** and **r'**.

# Figure 4

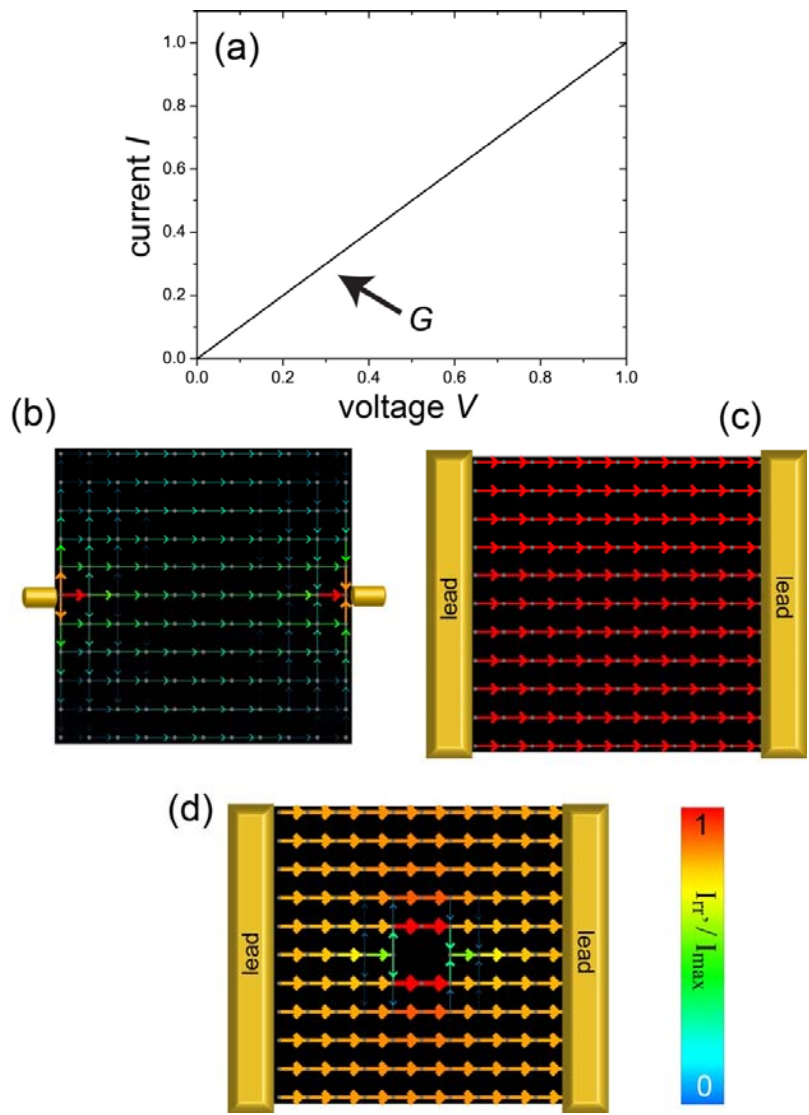

(a) Linear *IV*-curve of a classical resistor.

(b) Spatial current pattern in a classical resistor network attached to narrow leads.

(c) Spatial current pattern in a classical resistor network attached to wide leads.

(b) Spatial current pattern in a classical resistor network attached to wide leads with a defect in the center.

Color (see legend) and thickness of the arrows represent the magnitude of the normalized current $I_{\mathbf{rr'}}/I_{max}$ (normalization occurs for each figure separately).

**Figure 5**

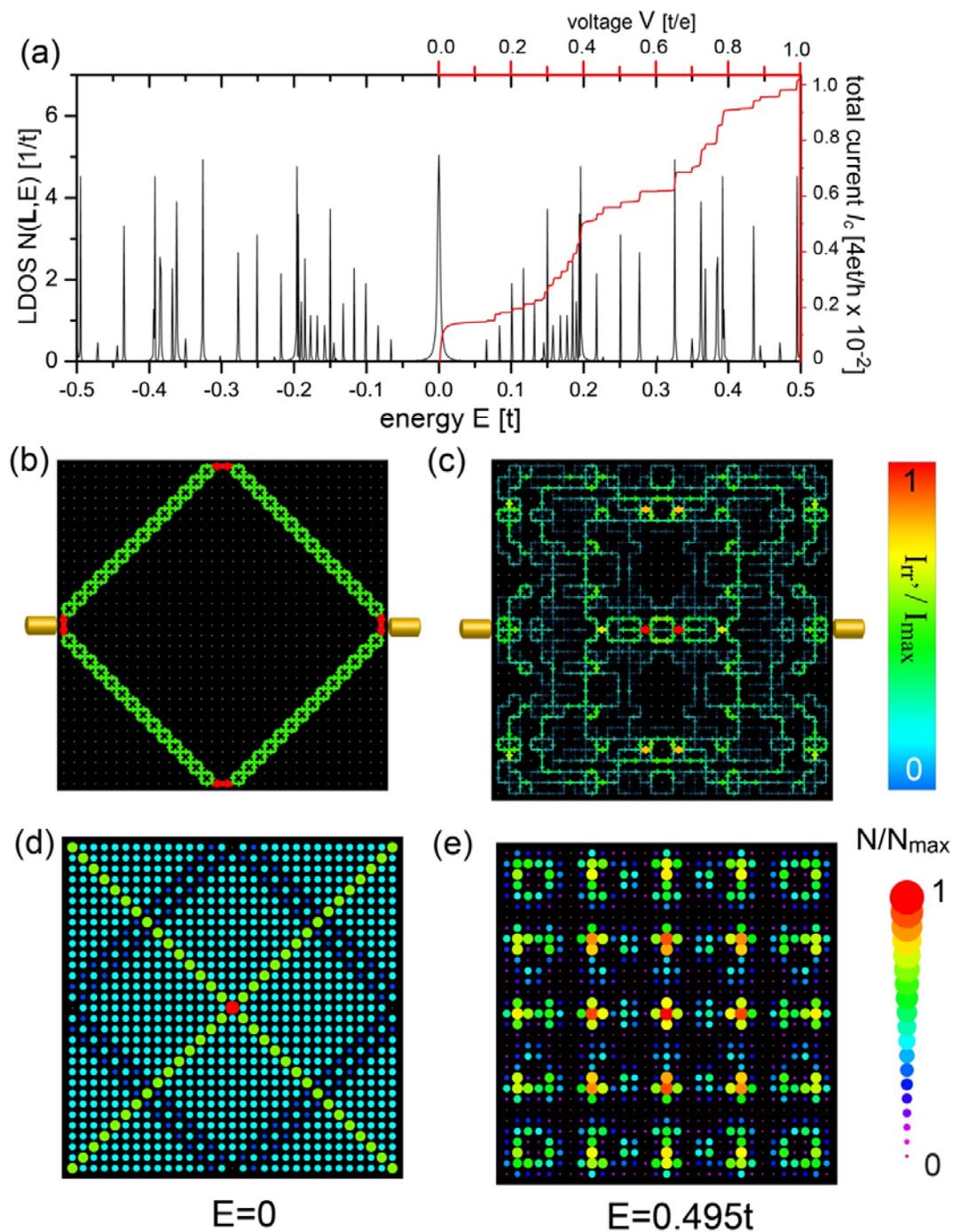

(a) Local density of states $N(\mathbf{L}, E)$ at $\mathbf{L}$ for a quantum network with $N_{x,y} = 31$. On the right hand side is overlain the total current $I_c$ as a function of applied voltage $V$.

(b) Spatial current pattern, $I_{rr'}$, carried by the $E_0 = 0$ state.

(c) Spatial current pattern, $I_{rr'}$, carried by the $E_1 = 0.495t$ state. This state is accessed for current transport by applying a gate voltage $V_g = E_1/e$ to the network.

(d) Local density of states $N(\mathbf{r}, E)$ at $E_0 = 0$.

(e) Local density of states $N(\mathbf{r}, E)$ at $E_1 = 0.495t$.

Color (see legend) and thickness of the arrows represent the magnitude of the normalized current $I_{rr'}/I_{max}$ (normalization occurs for each figure separately).

# Figure 6

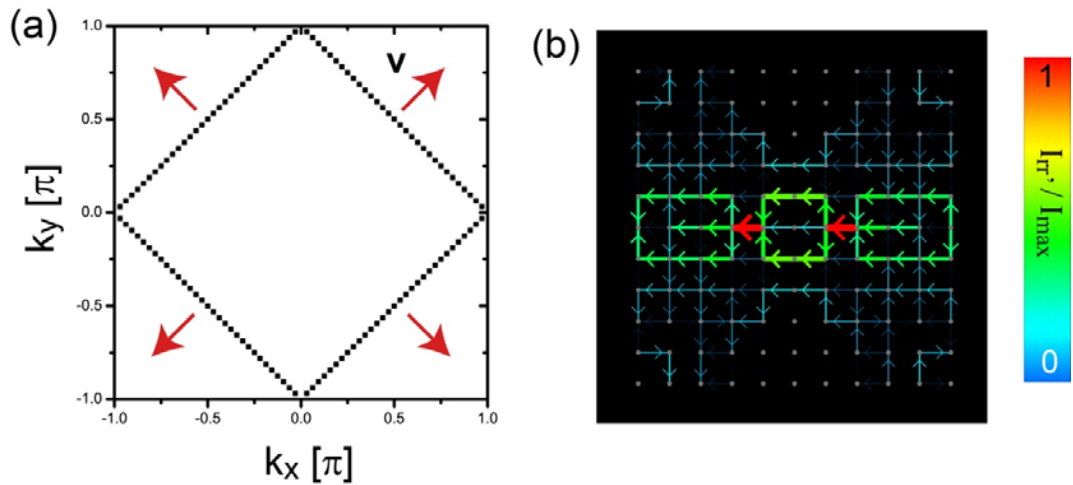

(a) States in the Brilluoin zone with $E_\mathbf{k} = 0$.

(a) Zoom-in of the spatial current pattern in the center of Fig.5(c), demonstrating the backflow of current (right to left) opposite to the applied voltage.

Color (see legend) and thickness of the arrows represent the magnitude of the normalized current $I_{\mathbf{rr}'}/I_{max}$ (normalization occurs for each figure separately).

# Figure 7

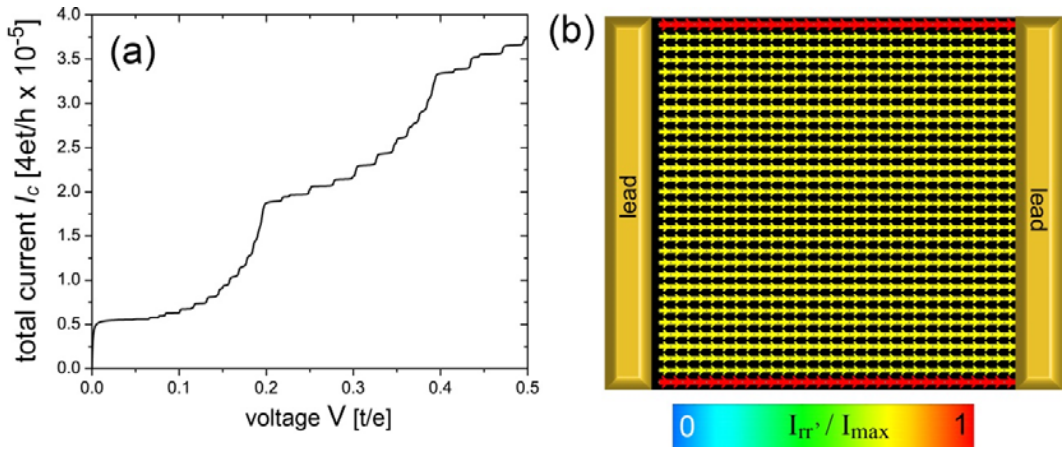

(a) Current as a function of applied voltage, $IV$-curve, for a network attached to wide leads.

(b) Spatial current pattern of the $E = 0$ state.

Color (see legend) and thickness of the arrows represent the magnitude of the normalized current $I_{rr'}/I_{max}$ (normalization occurs for each figure separately).

**Figure 8**

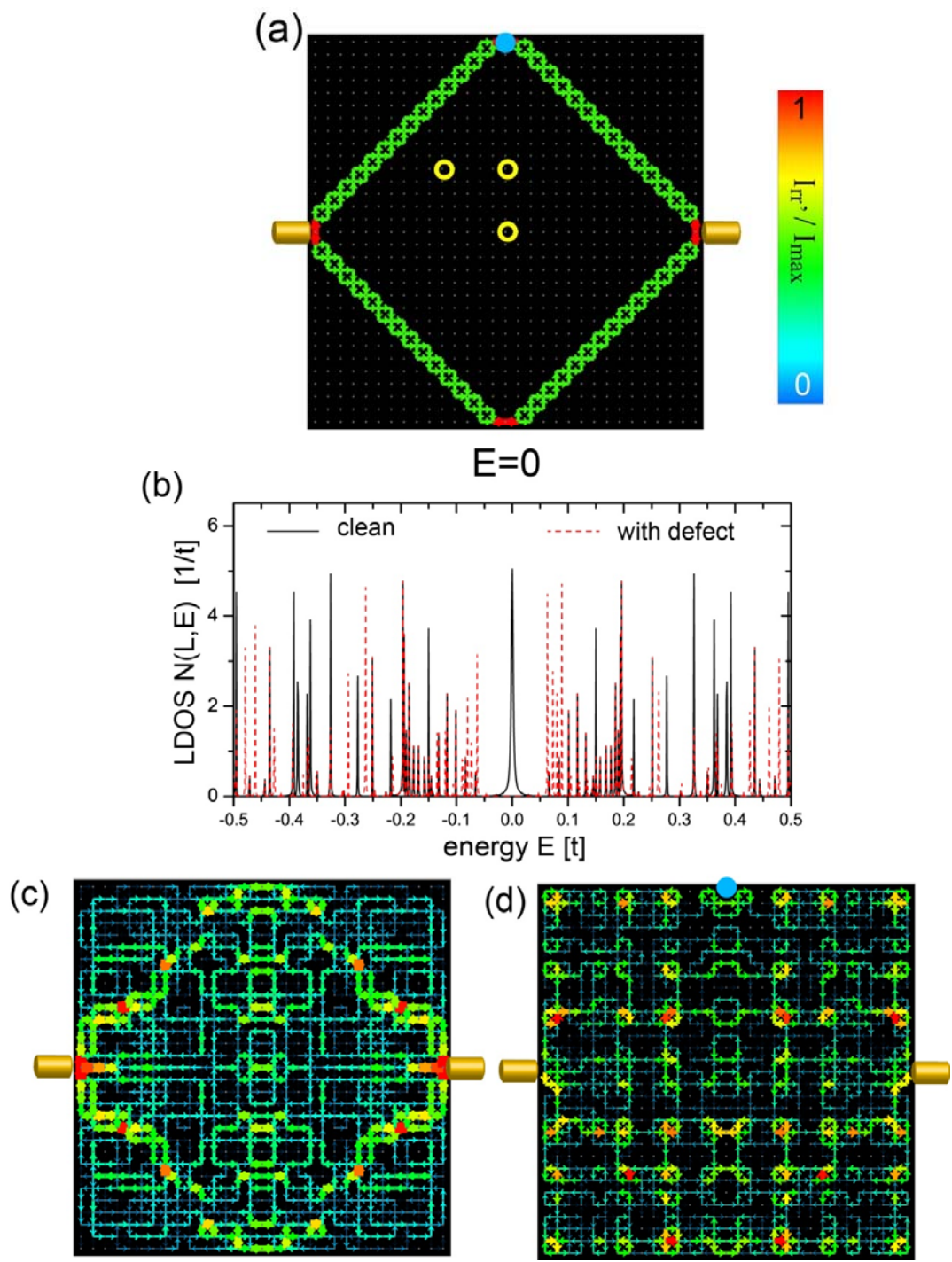

(a) Spatial current pattern, $I_{rr'}$, carried by the $E_0 = 0$ state in the clean network. Possible defect locations are indicated by circles.

(b) Comparison of the local density of states $N(\mathbf{L}, E)$ at $\mathbf{L}$ in quantum networks with and without a defect. The defect location is indicated by the filled blue circle in (a).

(c) Spatial current pattern, $I_{rr'}$, carried by the $E_1 = 0.435t$ state in the clean network. This state is accessed for current transport by applying a gate voltage $V_g = E_1/e$ to the network.

(d) Spatial current pattern, $I_{rr'}$, carried by the $E_1 = 0.435t$ state in the network containing a defect. The defect location is indicated by the filled blue circle. This state is accessed for current transport by applying a gate voltage $V_g = E_1/e$ to the network.

Color (see legend) and thickness of the arrows represent the magnitude of the normalized current $I_{rr'}/I_{max}$ (normalization occurs for each figure separately).

# Figure 9

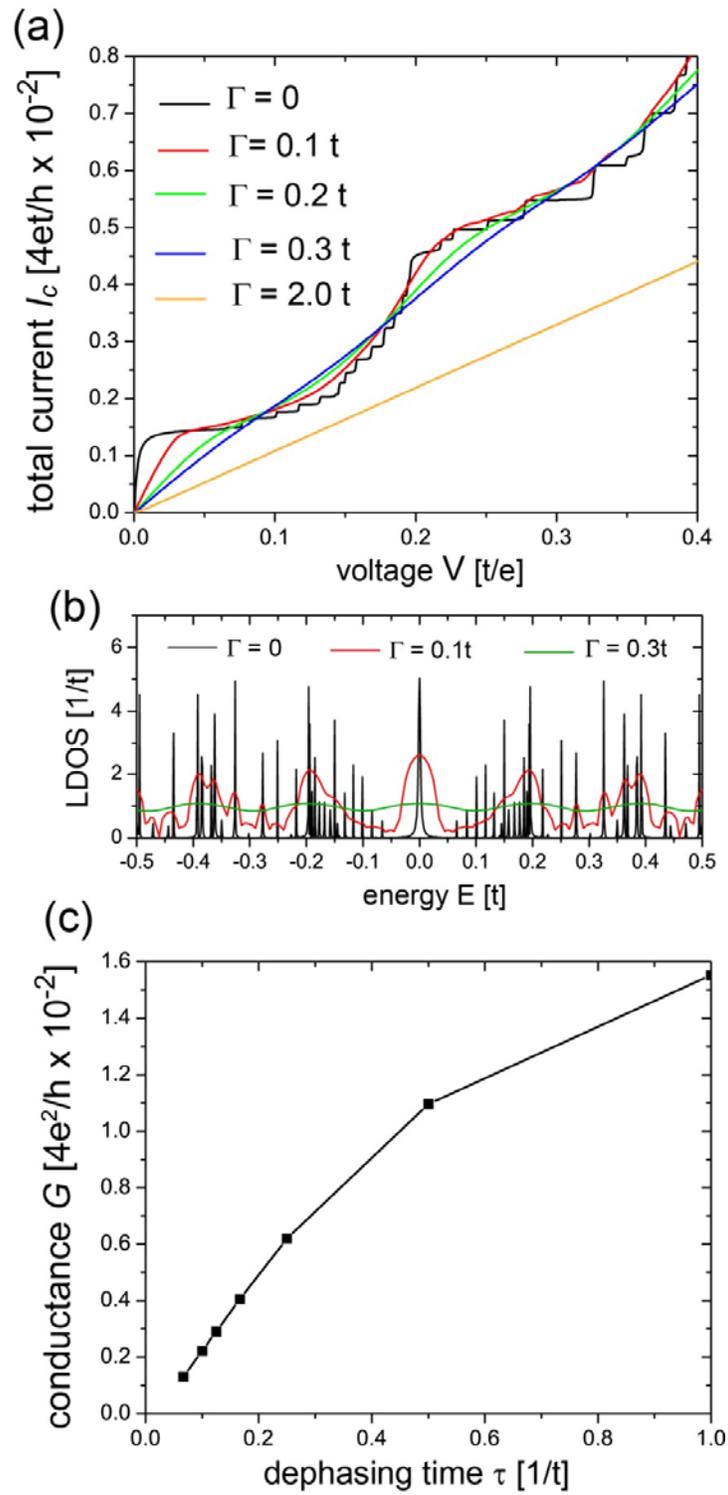

(a) *IV*-curves for different dephasing rates Γ.

(b) Comparison of the local density of states $N(\mathbf{L}, E)$ at $\mathbf{L}$ for different dephasing rates $\Gamma$

(a) Conductance $G$ as a function of dephasing time (or lifetime) $\tau$.

# Figure 10

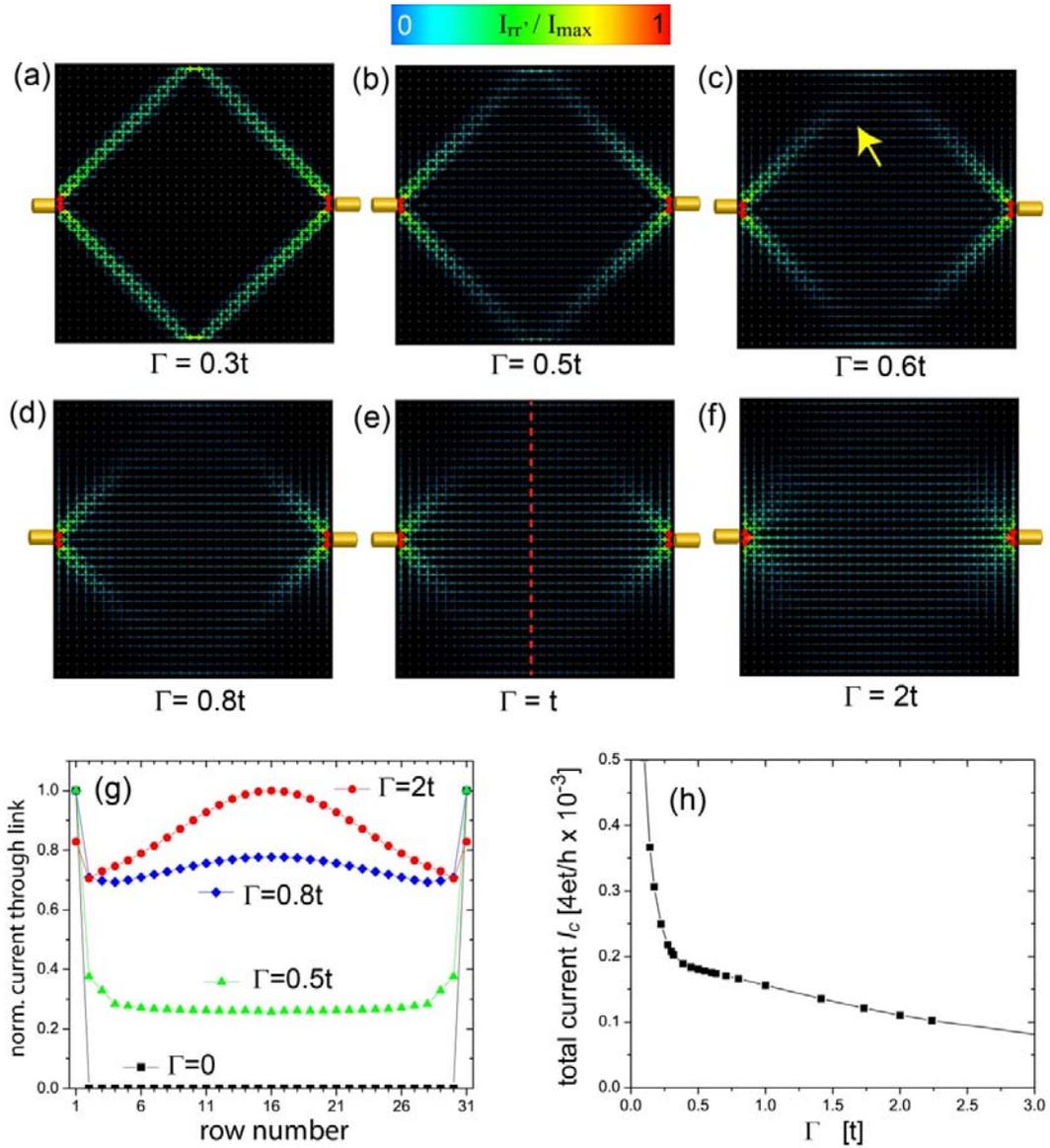

(a) – (f) Spatial current patterns, $I_{rr'}$ carried by the $E_0 = 0$ state for different dephasing rates $\Gamma$.

(g) Normalized current flowing through horizontal links in the center column of the network, as indicated by a dashed red line in (e) for different dephasing rates $\Gamma$.

(h) Total current $I_c$ carried by the $E_0 = 0$ state as a function of dephasing rate $\Gamma$.

Color (see legend) and thickness of the arrows represent the magnitude of the normalized current $I_{rr'}/I_{max}$ (normalization occurs for each figure separately).

# Figure 11

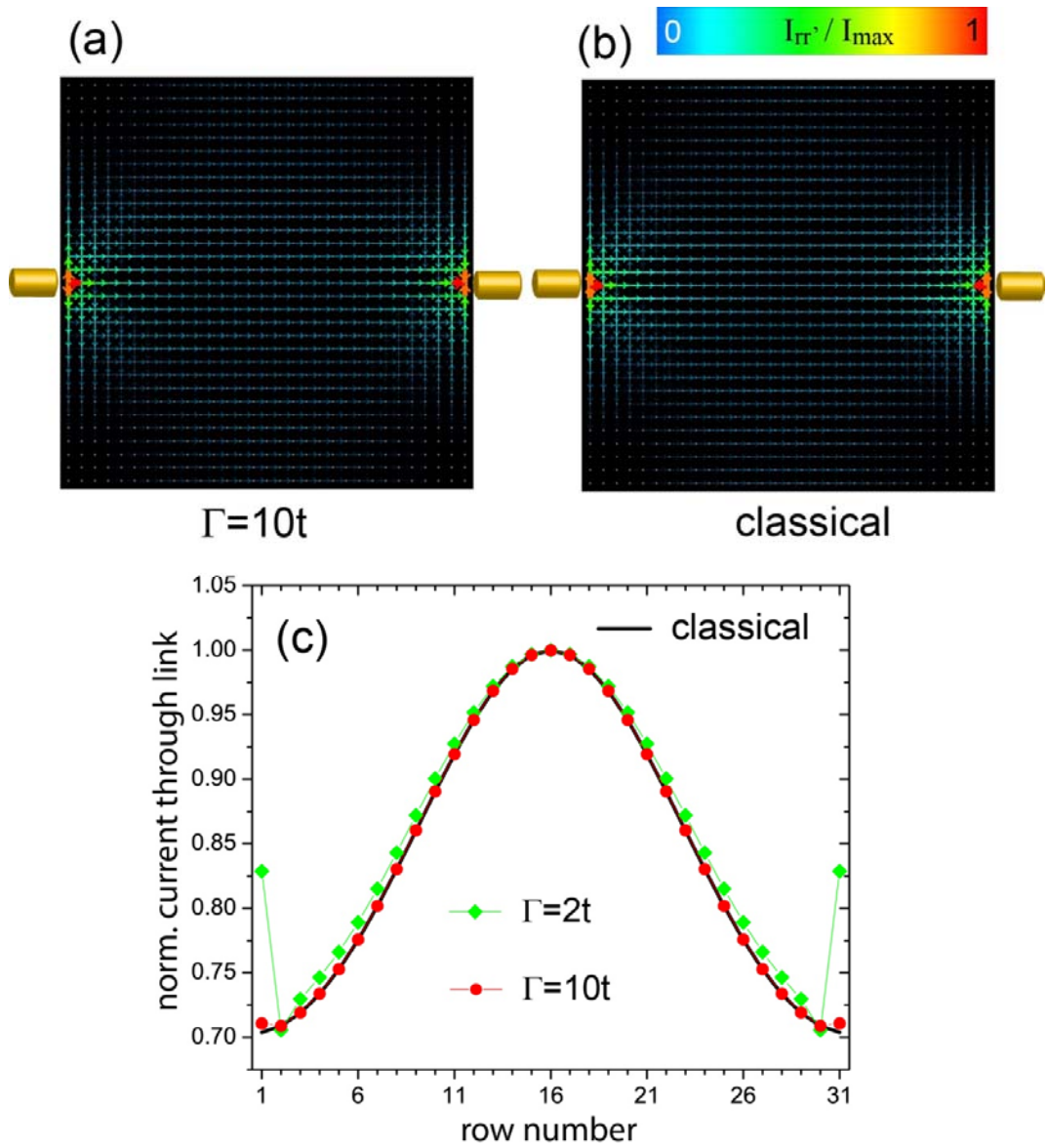

(a) Spatial current pattern, $I_{rr'}$ carried by the $E_0 = 0$ state for $\Gamma = 10t$.

(b) Spatial current pattern, $I_{rr'}$ in a classical resistor network.

(c) Normalized current flowing through horizontal links in the center column of the network, as indicated by a dashed red line in 10(e) for networks with different dephasing rates $\Gamma$, and the classical resistor network.

# Figure 12

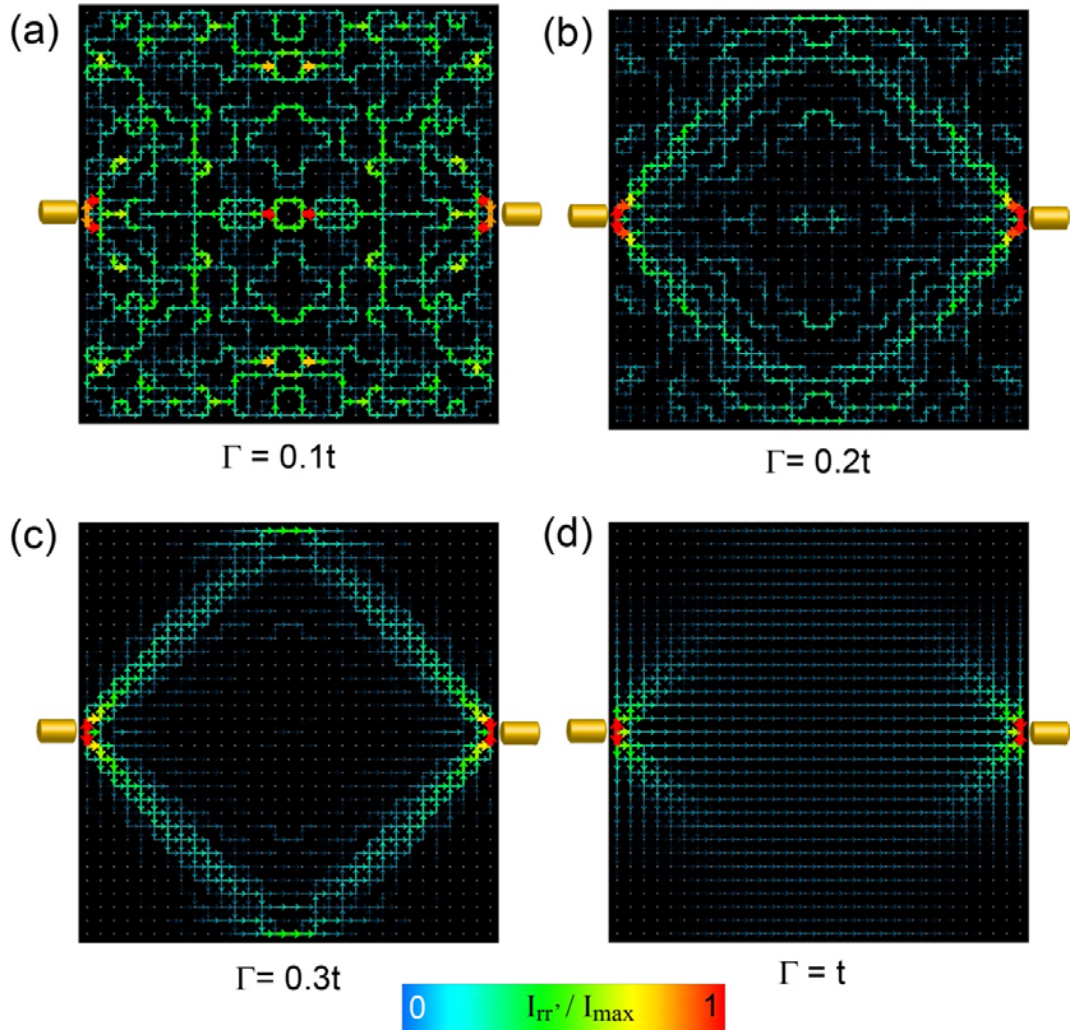

(a) – (d) Spatial current pattern, $I_{rr'}$, carried by the $E_1 = 0.495t$ state for different dephasing rates $\Gamma$. This state is accessed for current transport by applying a gate voltage $V_g = E_1/e$ to the network.

Color (see legend) and thickness of the arrows represent the magnitude of the normalized current $I_{rr'}/I_{max}$ (normalization occurs for each figure separately).

# Figure 13

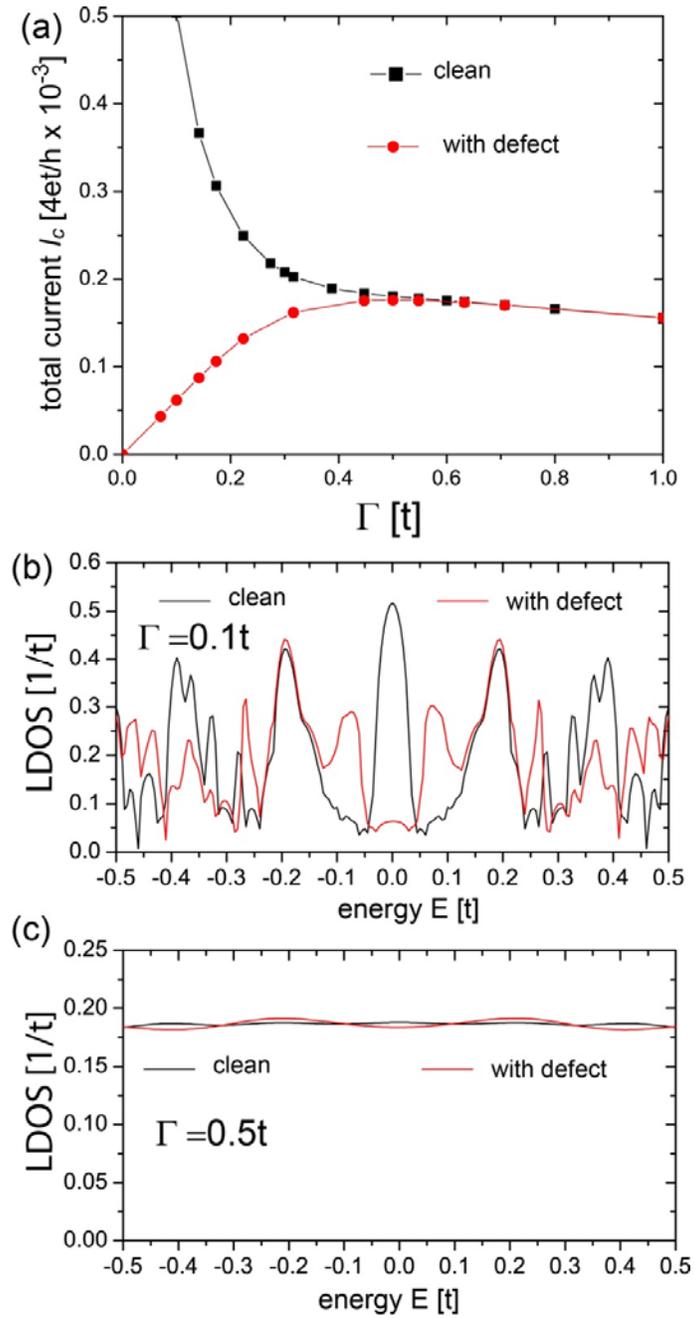

(a) Total current $I_c$ carried by the $E_0 = 0$ state as a function of dephasing rate $\Gamma$ for a network with and without a defect.

(b) Local density of states $N(\mathbf{L}, E)$ at $\mathbf{L}$ for a network with and without a defect. for $\Gamma = 0.1t$.

(c) Local density of states $N(\mathbf{L}, E)$ at $\mathbf{L}$ for a network with and without a defect. for $\Gamma = 0.5t$.

# Figure 14

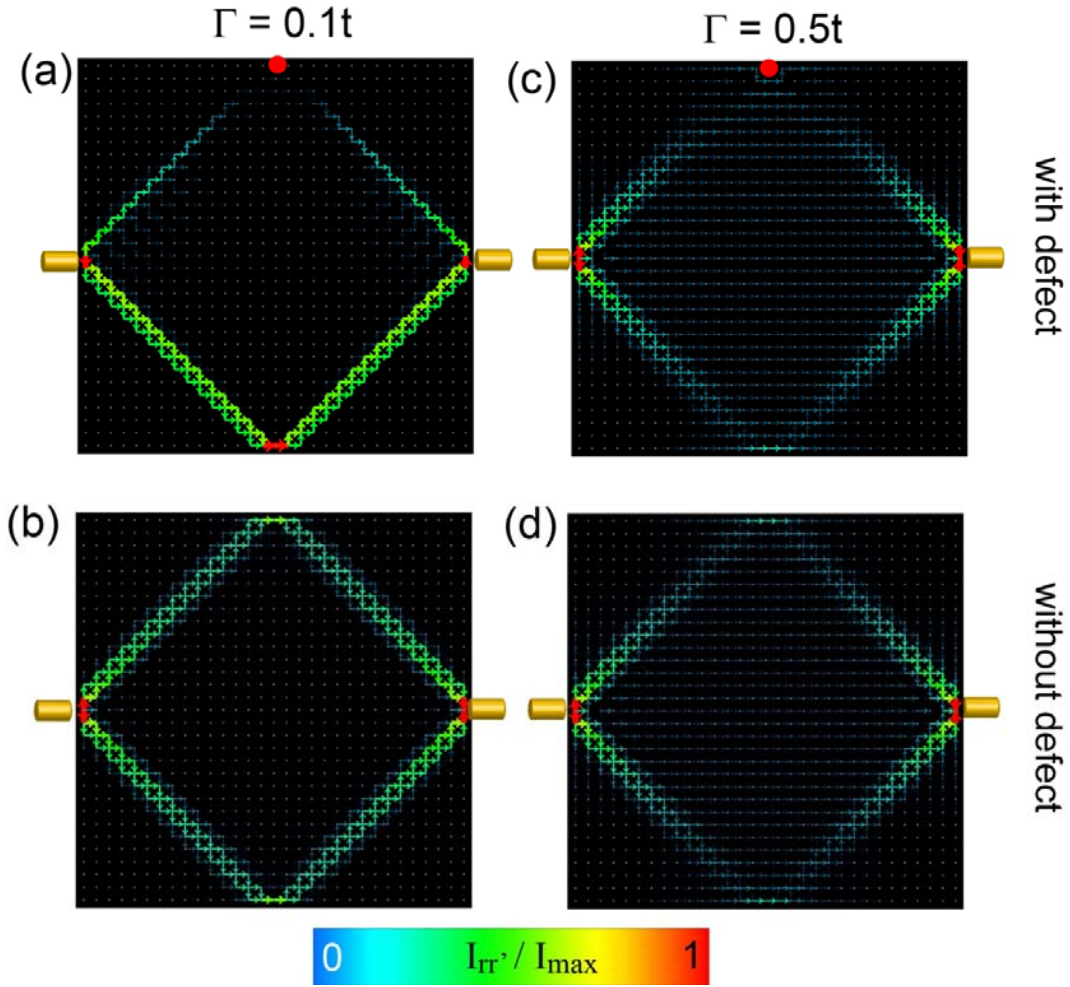

(a), (c) Spatial current pattern, $I_{rr'}$ carried by the $E_0 = 0$ state for different dephasing rates $\Gamma$ in the presence of a defect.

(b), (d) Spatial current pattern, $I_{rr'}$ carried by the $E_0 = 0$ state for different dephasing rates $\Gamma$ in the clean network.

Color (see legend) and thickness of the arrows represent the magnitude of the normalized current $I_{rr'}/I_{max}$ (normalization occurs for each figure separately).

## Figure 15

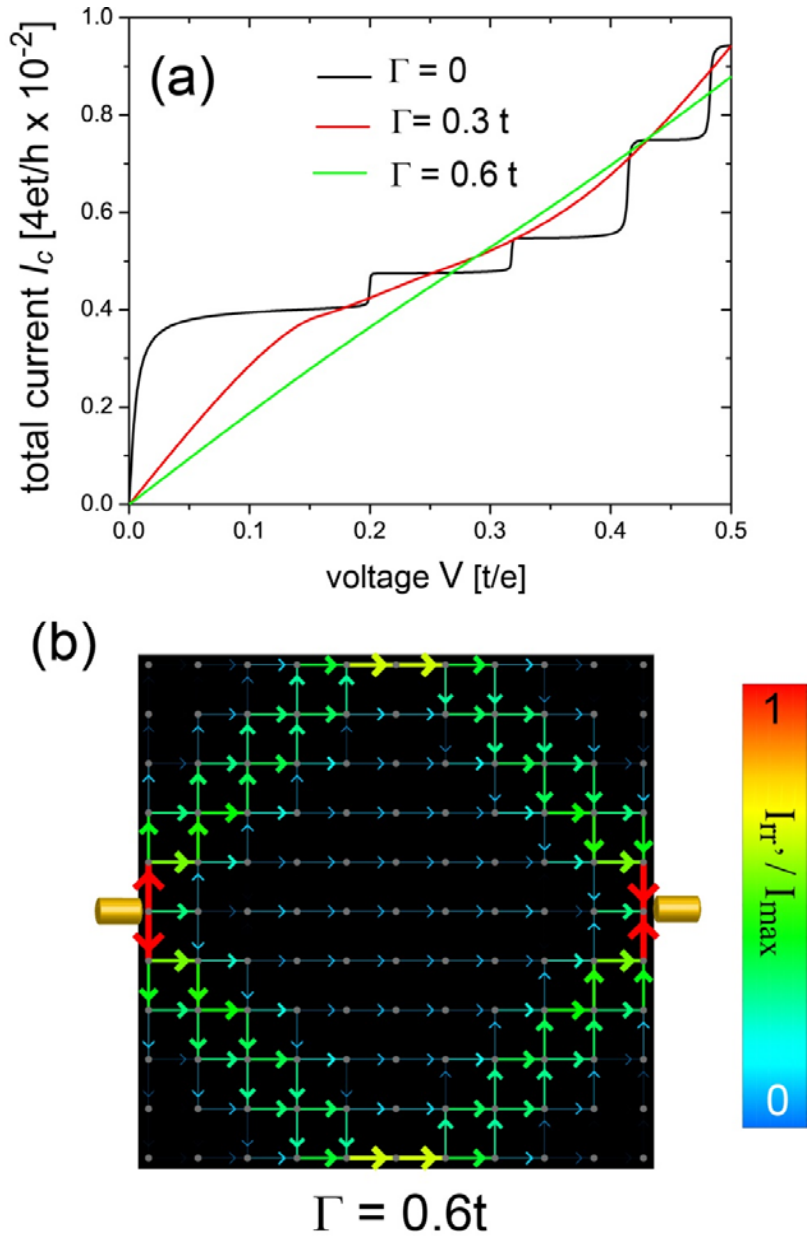

(a) *IV*-curves of a network with $N_{x,y} = 11$ for different dephasing rates $\Gamma$.

(b) Spatial current pattern, $I_{rr'}$ carried by the $E_0 = 0$ state for $\Gamma = 0.6t$.

Color (see legend) and thickness of the arrows represent the magnitude of the normalized current $I_{rr'}/I_{max}$ (normalization occurs for each figure separately).

# Figure 16

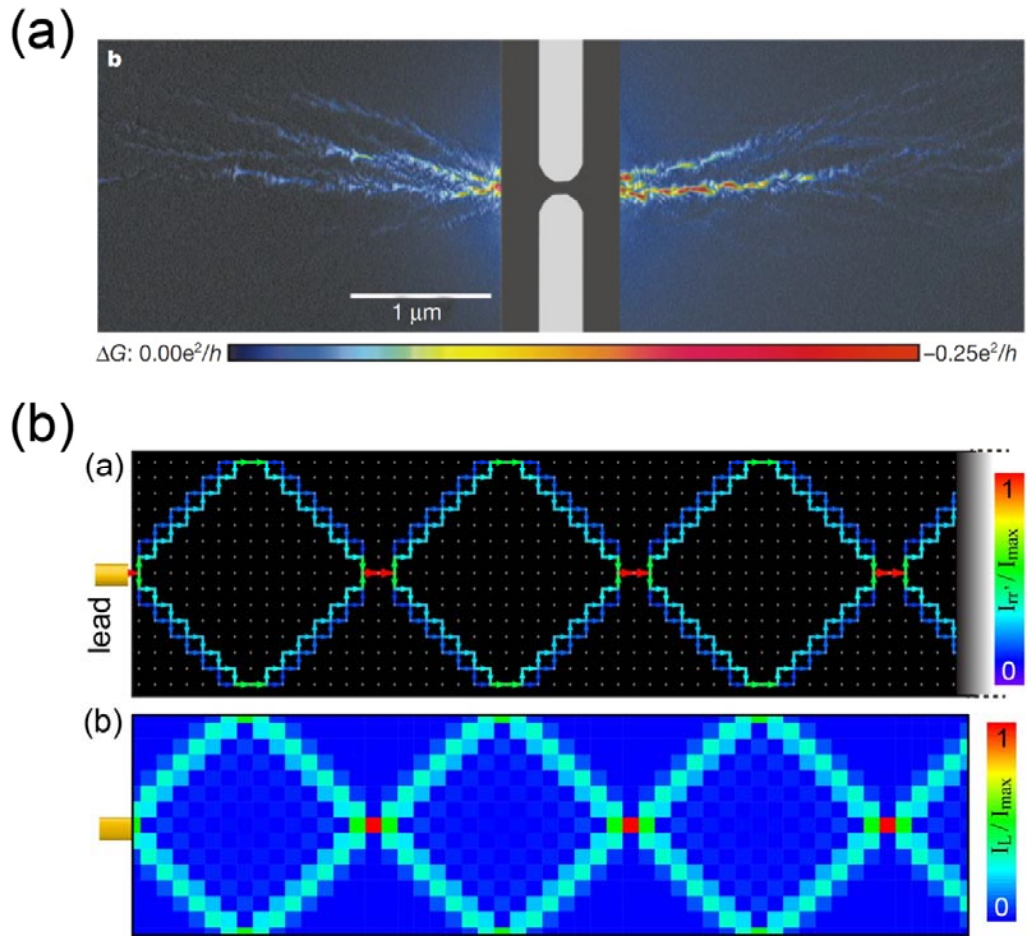

(a) Imaging of spatial current patterns in a two-dimensional electron gas near a quantum point contact using a scanning probe microscope (figure taken from Ref.[60])

(b) Imaging of spatial current patterns in a nanoscale network using a scanning tunnelling microscope (figure taken from Ref.[10]). Upper panel: actual spatial current pattern. Lower panel: imaged current pattern.

# Figure 17

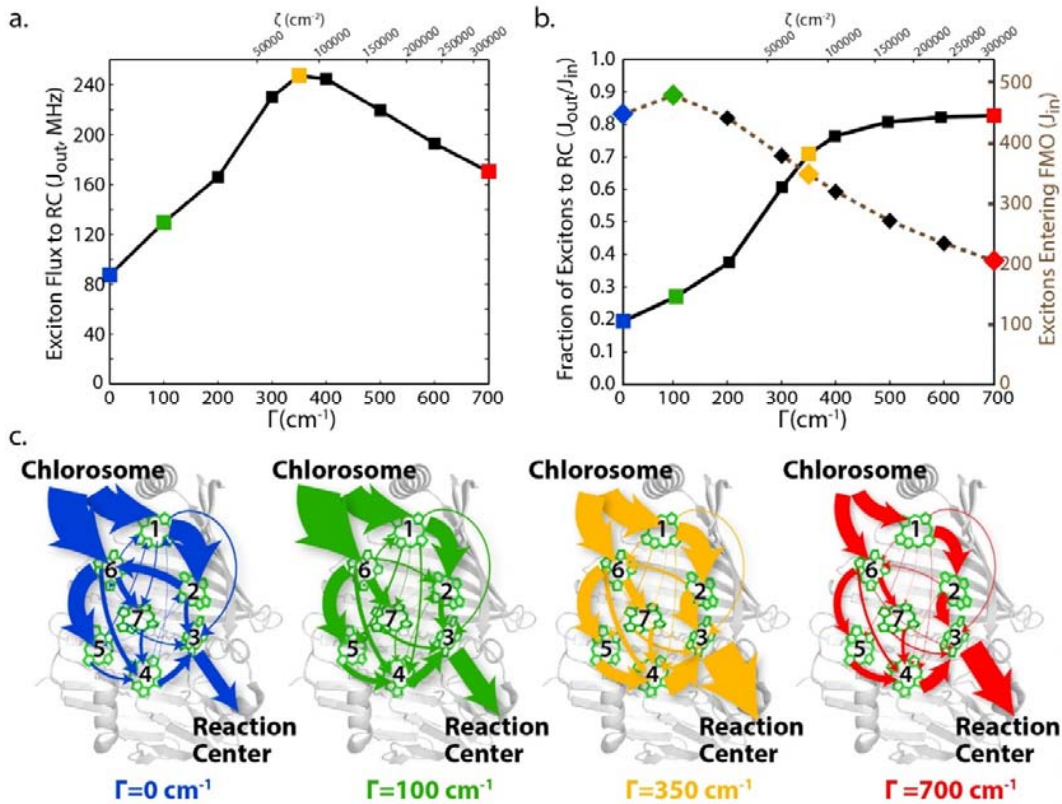

(a) Excitonic current, $I_{out}$, from FMO to the reaction center as a function of dephasing rate $\Gamma$.

(b) Excitonic current, $I_{in}$, from the chlorosome into the FMO dashed line, and transport efficiency $I_{out}$, /$I_{in}$ as a function of dephasing rate $\Gamma$.

(c) The spatial excitonic current pattern between chromophore sites in FMO for several values of the dephasing rate. Thickness of arrows is linearly proportional to the magnitude of the current. Figure is taken from Ref.[33]

# Figure 18

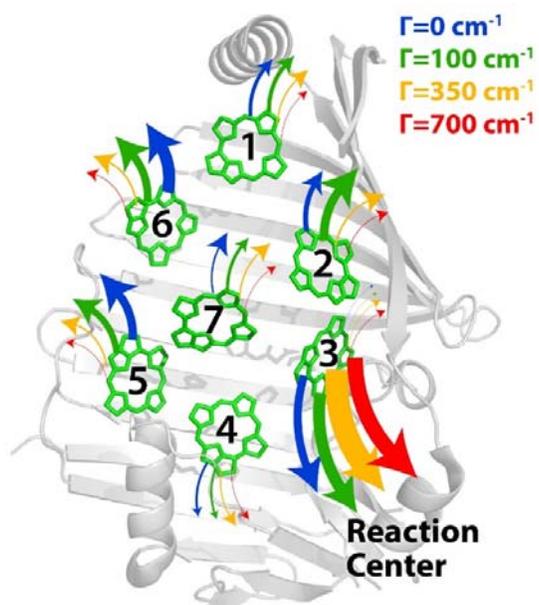

Shown are the exciton recombination rates at the seven chromophore sites and the exciton current from FMO (site 3) to the reaction center for various dephasing rate. The thickness of the arrows is proportional to the recombination rate at each site. Figure is taken from Ref.[33].